\documentclass[twocolumn]{aastex631}
\usepackage{bm}

\usepackage{hyperref}
\usepackage{bbding}
\usepackage{tabularx}

\shorttitle{}
\shortauthors{Qin et al.}

\begin{document}

\title{Active Galactic Nuclei and STaR fOrmation in Nearby Galaxies (AGNSTRONG). II: Results for Jetted Type-I AGNs with Strong Ionized Gas Outflows}

\author[0009-0000-0126-8701]{Chen Qin\textsuperscript{\Envelope}}\email{qinc@mail.ustc.edu.cn}
\author[0000-0003-1270-9802]{Huynh Anh N. Le\textsuperscript{\Envelope}}\email{lha@ustc.edu.cn}
\author[0000-0002-1935-8104]{Yongquan Xue\textsuperscript{\Envelope}}\email{xuey@ustc.edu.cn}
\author[0000-0002-1653-4969]{Shifu Zhu}
\author[0000-0000-0000-0000]{Xiaozhi Lin}
\affiliation{Department of Astronomy, University of Science and Technology of China, Hefei 230026, China}
\affiliation{School of Astronomy and Space Science, University of Science and Technology of China, Hefei 230026, China}
\author[0000-0000-0000-0000]{Kim Ngan Nhat Nguyen}
\affiliation{Faculty of Physics and Engineering Physics, University of Science, Ho Chi Minh City, Vietnam}
\affiliation{Vietnam National University, Ho Chi Minh City, Vietnam}

\begin{abstract}

We investigate the correlation between ionized gas outflows, jets, and star formation in a sample of 42 local type-I active galactic nuclei (AGNs) exhibiting significant [O III] outflows. This study uses both new submillimeter (sub-mm) observations and archival data from the James Clerk Maxwell Telescope. Our analysis, which includes a correction for jet emission in the sub-mm bands, fitting spectral energy distribution, and analyzing spectra, enables us to derive star-formation rates (SFRs) through various methods. By comparing radio power and SFRs, we select a sub-sample of jetted AGNs of which radio emission is mostly from the jets. We find that jetted AGNs predominantly lie above the main sequence of star-forming galaxies, suggesting a correlation between jet activity and star formation. By comparing dust extinction, we demonstrate that jetted AGNs do not have more dust which is the fuel of both star formation and AGN activity. Therefore, this correlation is more likely to arise from AGN feedback. We also find that the Eddington ratio does not impact the specific SFRs (sSFRs) of our sample. Additionally, for jetted AGNs, stronger radio emission corresponds to higher sSFRs, suggesting that jet emission may promote star formation, i.e., positive feedback. Our results not only shed light on the feedback mechanisms of AGNs but also underscore the complex interplay between black hole activity and star formation in galaxy evolution.

\end{abstract}

\keywords{\href{http://astrothesaurus.org/uat/16}{Active galactic nuclei (16)} --- \href{http://astrothesaurus.org/uat/1569}{Star formation (1569)} --- \href{http://astrothesaurus.org/uat/2129}{Spectral energy distribution (2129)}}

\section{Introduction} \label{sec:intro}

Active galactic nuclei (AGNs) are supermassive black holes (SMBHs) at the centers of galaxies that are accreting matter. Previous research found significant correlations between the properties of SMBHs and their host galaxies, such as the well-known correlation between the mass of SMBHs and that of the galactic bulges. The correlations suggest a co-evolution between SMBHs and their host galaxies, which might be regulated by AGN feedback (e.g., \citealt{kormendy_coevolution_2013, xue_chandra_2017}). However, the details of how AGN activity connects to star formation in host galaxies and the impact of AGN energy outputs, such as ionized gas outflows, on the interstellar medium (ISM) remains unclear. Negative and positive feedback are the two main scenarios for explaining the observational results \citep{le_medium_2014, le_ionized-gas_2017, zubovas_agn_2013}. The role of AGN outflows, whether they suppress star formation by expelling/heating the ISM or trigger starburst by compressing it, complicates the galaxy evolution.

A key challenge in investigating AGN feedback is precisely measuring the star-formation rates (SFRs). SFR estimations suffer from considerable uncertainties ($\sim$0.7 dex, \citealt{merloni_cosmic_2010}), arising from the systematics of various SFR indicators such as ultraviolet (UV), optical, or infrared (IR) measurements. The UV approach measures SFR from young stars but struggles with dust obscuration \citep{salim_uv_2007}. Optical emission lines (e.g., [O II] $\lambda$3727, H${\rm \alpha}$, \citealt{zhuang_recalibration_2019}), even after correcting for dust, are muddied by contributions from young stars, ISM shocks, or AGNs. Mid-infrared (MIR) signals may also include AGN influences. Far-infrared (FIR) or submillimeter (sub-mm) diagnostic captures cold dust emission mostly heated by star formation, with minimal interference from ISM shocks and dusty torus AGN emission (e.g., \citealt{ellison_infrared_2016,xu_agns_2020}). Spectral energy distribution (SED) fitting can decompose stellar-heated dust emission and present a reliable SFR estimation, in which FIR and sub-mm observations are still crucial to restrict dust models. \cite{kim_determining_2022} demonstrated that excluding FIR/sub-mm fluxes can lead to overestimation of SFRs by a factor of two.

In the unified models of AGNs, the distinction between type-I and type-II AGNs mainly results from orientation with the obscuring dusty torus. Type-I AGNs show unobscured broad-line emission, while type-II AGNs are viewed through the dusty torus and lack visible broad lines \citep{1993ARA&A..31..473A, 1995PASP..107..803U}. Because type-I AGNs provide direct views of the broad-line region and the optical continuum, they allow reliable single-epoch measurements of black hole masses and accretion rates (e.g., \citealt{2013BASI...41...61S}).

Some observations found positive correlations between AGN activity and star formation in their host galaxies. \cite{woo_correlation_2020} found type-II AGNs with larger strength of outflows tend to have higher specific star-formation rates (sSFRs). \cite{ayubinia_investigating_2023} found that AGNs with higher radio-Eddington ratios (the radio luminosity divided by Eddington luminosity) systematically exhibited higher [O III] velocity dispersion across the entire dynamic range of [O III] luminosity or Eddington ratio, implying a connection between ionized gas outflows and large-scale radio jets. Therefore, to fully understand AGN feedback, it is necessary to consider the effects of gas outflows and jets on SFRs simultaneously. We have assembled a sample of 42 type-I AGNs with strong ionized gas outflows. Their spectra exhibit broad H${\rm \beta}$ emission lines, allowing for the estimation of SMBH mass and Eddington ratios using the single-epoch method. To robustly measure the SFRs in these AGN hosts, FIR and sub-mm data from JCMT, together with multiband data, are important for constraining the dust emission and the SFRs in SED fitting. Moreover, radio observations provide a probe of jet power, allowing us to directly examine the connection between jet activity, ionized outflows, and star formation.

This paper is the second paper in the AGNSTRONG series, which investigates the interactions between AGNs and their host galaxies (for details, see \citealt{le_active_2024}). \cite{xia_active_2024}, the third paper in this series, utilizes spatially resolved spectroscopic observations along the slit direction to analyze the properties of ionized and molecular gas for six local ($z<0.1$) AGNs.

In this work, we used the Submillimetre Common-User Bolometer Array 2 (SCUBA-2, \citealt{holland_scuba-2_2013}) on the James Clerk Maxwell Telescope (JCMT) to observe 21 radio-luminous AGNs. We utilized archival SCUBA-2 data for another 21 AGNs with moderate radio luminosity. We employed various methodologies to measure the SFRs of these AGNs. We aimed to investigate the correlation among jets, ionized gas outflows, and SFRs, thereby shedding light on AGN feedback. In Section~\ref{sec:2Observations}, we outline the sample selection, JCMT observation, the reduction of sub-mm data, and the collection of multiband data. In Section~\ref{sec:3Methodology} we elaborate on the analyses for radio data, SED, and spectral fitting. In Section~\ref{sec:4Results} we present the main results and discussions. Further, A summary of this paper is presented in Section~\ref{sec:6Summary}. Following  \cite{kim_determining_2022}, we adopt the stellar initial mass function (IMF) from \cite{chabrier_galactic_2003}. For comparison, the SFRs using the \cite{kroupa_variation_2001} and \cite{salpeter_luminosity_1955} IMFs should be adjusted by a multiplication factor of 0.94 and 0.6 \citep{madau_cosmic_2014}. Throughout this paper, we adopt the $\mathrm{\Lambda CDM}$ cosmology with $H_0=\mathrm{70\ km\ s^{-1}\ Mpc^{-1}}$, ${\rm \Omega_M=0.3}$, and ${\rm \Omega_\Lambda=0.7}$. 

\section{Sample selection, JCMT observations, and Multiband data} \label{sec:2Observations}

\subsection{Sample Selection} \label{subsec:SampleSelection}

\begin{figure}[t]
\includegraphics[width=1.0\linewidth]{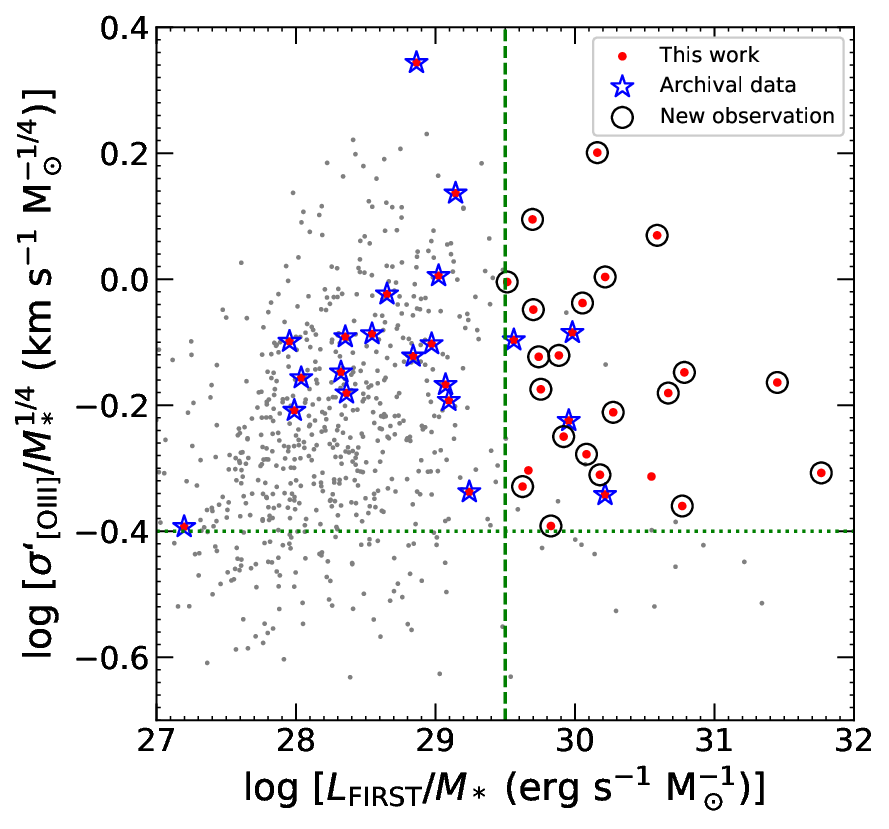}
\caption{Distribution of $\log\ \sigma_{\rm [OIII]}'/M_*^{1/4}$ as a function of radio luminosity normalized by stellar mass (from \href{https://www.sdss3.org/dr10/spectro/galaxy_mpajhu.php}{MPA--JHU}). Here, $M_*^{1/4}$ serves as a proxy for stellar velocity dispersion. Gray dots represent our parent sample of AGNs with $z<0.3$ \citep{rakshit_census_2018}. The red points our sample of 42 type-I AGNs investigated in this study, while the black circles highlight objects with new SCUBA-2 observations. Blue stars are targets for which archival SCUBA-2 data are available. The green dashed line and dotted line show thresholds of $L_{\rm 1.4GHz}/M_{*}{\rm >10^{29.5}\ erg\ s^{-1}\ M_{\odot}^{-1} }$ and ${\rm log}\ \sigma'_{\rm [OIII]}/M_*^{1/4}{\rm >-0.4\ km\ s^{-1}\ M_{\odot}^{-1/4}}$, indicative of objects with high radio luminosity and strong ionized gas outflows, respectively.
\label{fig:1Sample}}
\end{figure}

\begin{figure*}[ht!]
\includegraphics[width=1.0\linewidth]{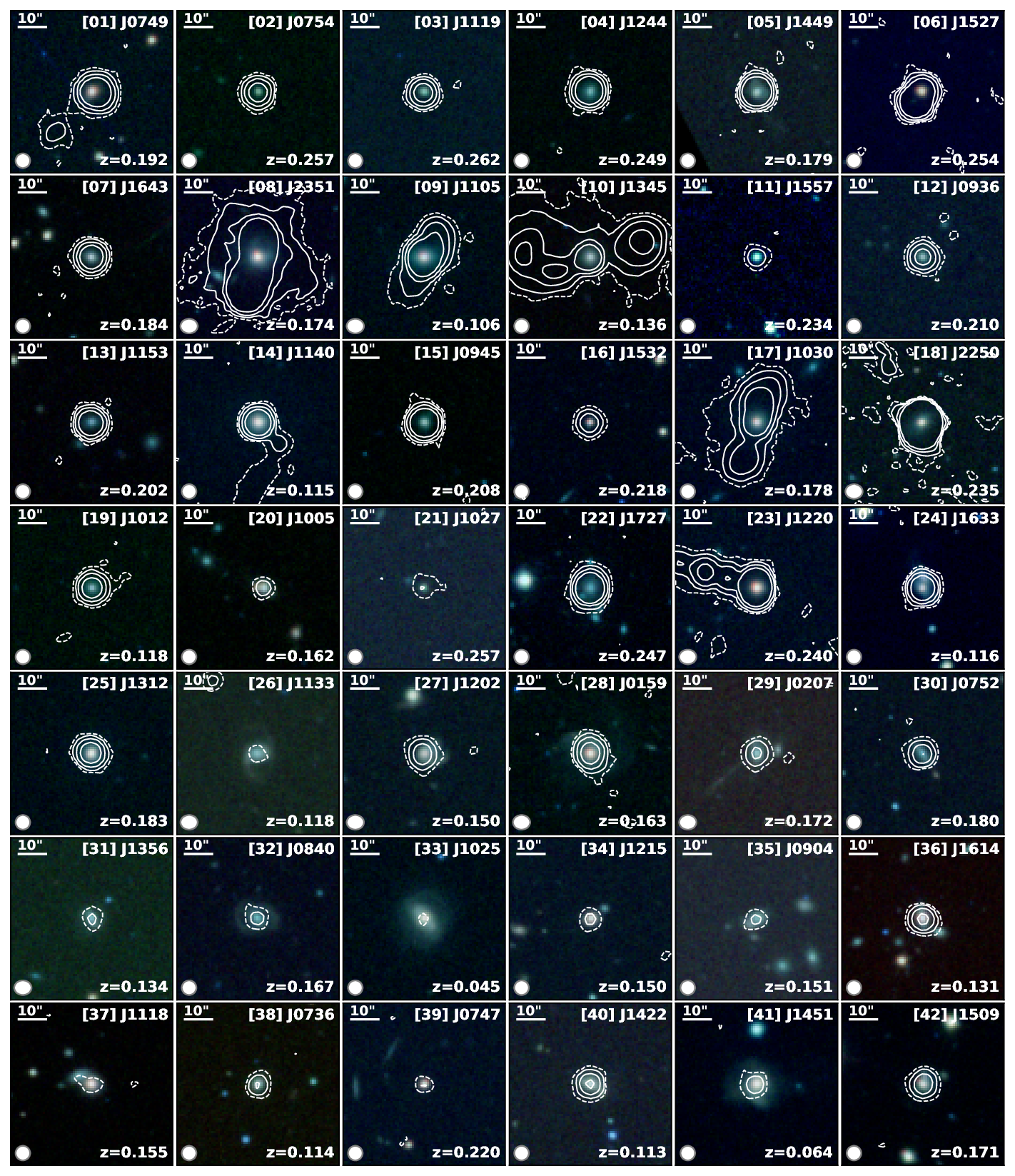}
\centering
\caption{
Optical $gri$-composite images for our sample of 42 objects from the SDSS database, with overlays of FIRST contours illustrating levels at $\rm 3~rms$ (dashed lines), $\rm 9~rms$, $\rm 27~rms$, and $\rm 81~rms$ (all solid lines) at $\mathrm{1.4~GHz}$. The IDs and names of the targets, positioned in the upper-right corner of each panel, correspond directly to those listed in Table \ref{tab:1ObsInfo}. The objects with IDs 1-21 are newly observed with SCUBA-2. The main beam sizes of FIRST are depicted in the lower-left corner of each panel. Additionally, the redshift of each target is displayed in the lower-right corner of each panel. The field of view of each panel is $\mathrm{80\ arcsec \times 80\ arcsec}$. The scale bar of 10~arcsec is shown in the upper-left corner of each panel.
\label{fig:2Images}}
\end{figure*}

\cite{rakshit_census_2018} conducted a sample of 5717 AGNs from the Sloan Digital Sky Survey (SDSS) DR12 catalog. They focused on objects classified as ``QSO" by the SDSS spectroscopic pipeline and chose the objects having $z<0.3$ and a signal-to-noise ratio (SNR) at $\rm 5100\ \AA>10$ and an amplitude-to-noise ratio of $\rm H\beta$ line being $>5$. They cross-matched their sample of AGNs with the Faint Images of the Radio Sky at Twenty-centimeters (FIRST) 1.4~GHz catalog \citep{helfand_last_2015} using a matching radius of 5 arcsec and found that 918 AGNs are detected in the radio band with a 1.4~GHz flux density limit of 1 mJy. The radio luminosities are estimated by the following formula:
\[L_{\rm FIRST} = \nu_R F_{\rm int} \times (1+z)  \times (4 \pi D_L^2), \]
where $\nu_R$ is the frequency of $\rm 1.4~GHz$, $F_{\rm int}$ is the integrated radio flux of the FIRST survey, and $D_L$ is the luminosity distance at the redshift of $z$. However, this estimation only considers the radio luminosity from the ``Core" component and does not perform a K-correction. We provide a more detailed calculation in Section~\ref{subsec:1.4GHz}.

The \cite{rakshit_census_2018} catalog is particularly well suited to the science goals of this work. Its strict SNR criteria ensure reliable measurements of continuum luminosities, broad-line widths, and black hole masses for all type-I AGNs, which are crucial for quantifying accretion properties. Most importantly, their systematic decomposition of the [O~III] $\lambda5007$ line yields robust indicators of ionized-gas outflows (e.g., velocity shift and dispersion), enabling a homogeneous characterization of AGN-driven kinematics across the sample. Combined with uniform FIRST cross-matching, this dataset provides a well-defined parent sample from which we can select our sample and investigate the connection between jet activity and star formation of host galaxy.

The 918 radio-detected AGNs are shown in Figure~\ref{fig:1Sample} as grey points. We select our sample from this catalog. Since our primary focus is on AGN feedback, i.e., the influence of AGN activity on their host galaxies, the radio luminosity normalized by the stellar mass of the host galaxy is a better tracer of feedback strength driven by the jet, compared to the commonly used radio-loudness parameter. Consequently, we employed ${L_{\rm FIRST}/M_{*}}$ to represent relative radio luminosity, where $M_*$ is the stellar masses from the \href{https://www.sdss3.org/dr10/spectro/galaxy_mpajhu.php}{MPA--JHU}, which is based on the methods of \cite{kauffmann_stellar_2003,2004MNRAS.351.1151B,2004ApJ...613..898T}.

In this study, we utilized $\sigma'^2_{\rm [OIII]}=\sigma^2_{\rm [OIII]}+V^2_{\rm [OIII]}$ as an indicator of the strength of ionized gas outflows, where $V_{\rm [OIII]}$ and $\sigma_{\rm [OIII]}$ are the first and second moments of ${\rm [O\ III]\ \lambda 5007}$ emission line. This formula takes into account the effects such as inclination, outflow geometry, and extinction, offering a more comprehensive description of the outflow kinematics (for details, see \citealt{bae_prevalence_2016}). Given that our objects are type-I AGNs, mostly characterized by spectra predominantly dominated by AGN radiation, precisely estimating the stellar velocity dispersion of their host galaxies presents a substantial challenge. Therefore, we employed $M_*^{1/4}$ as a proxy for stellar velocity dispersion to indicate the strength of the gravitational kinematic component, which circumvents the difficulties associated with direct measurements in these AGN-dominated systems. We have utilized the Fiber–Jackson relation following \cite{woo_delayed_2017}.

Among the 918 AGNs detected in the radio band, 21 objects were found to have publicly available SCUBA-2 data within 1 arcminute.\footnote{\url{https://proposals.eaobservatory.org/jcmt/tool/clash}} These are depicted as blue stars in Figure~\ref{fig:1Sample}. The majority of these objects have moderate $L_{\rm FIRST}/M_*$, as shown in Figure~\ref{fig:1Sample}. To extend the range of relative radio luminosity, we selected 21 AGNs with $L_{\rm FIRST}/M_{*} {\rm > 10^{29.5}\ erg\ s^{-1}\ M_{\odot}^{-1}}$. We also require the objects to have significant ionized outflows with ${\rm log}\ \sigma'_{\rm [OIII]}/M_*^{1/4} {\rm> -0.4\ km\ s^{-1}\ M_{\odot}^{-1/4}}$. The two criteria above are indicated by the green dashed and dotted lines in Figure ~\ref{fig:1Sample}, respectively. The 21 newly observed AGNs are marked by black circles.

Since \cite{rakshit_census_2018} selected samples using only redshift, SNR and classification of ``QSO" by the SDSS spectroscopic pipeline, there might be type-II AGNs mixed in. Therefore, In Subsection~\ref{subsec:34Spectra} we carefully fitted the spectra of our 42 objects and found that although some objects do not have significant broad $\rm H\beta$ emission lines, all objects have the full widths at half maximum (FWHMs) of $\rm H\alpha $ emission lines larger than $\rm 1200\ km/s$, and the majority (34/42) have FWHMs $\rm >2000\ km/s$, so we consider them all to be type-I AGNs. Furthermore, in Section~\ref{sec:4Results}, we demonstrated that the radio emission from some objects primarily comes from the jets. Following the suggestion of \cite{padovani_active_2017}, we refer to them as jetted type-I AGNs.

To summarize, we selected a sample of 42 nearby type-I AGNs with moderate and high relative radio luminosity and strong ionized outflows, comprised of 21 objects with new SCUBA-2 observations conducted by us and another 21 with archival data. The SDSS images with FIRST contours for these AGNs are illustrated in Figure~\ref{fig:2Images}, where the FIRST contours are depicted as white dashed and solid lines.

\begin{deluxetable*}{cccccccccccc}[ht!]
\tabletypesize{\scriptsize}
\tablewidth{0pt} 
\tablecaption{Observation Information \label{tab:1ObsInfo}}
\setlength{\tabcolsep}{1.8mm}{}
\tablehead{
\colhead{ID} & \colhead{Name} & \colhead{R.A.} & \colhead{Dec.} & \colhead{$z$} & \colhead{Obs.} & \colhead{$T_{\rm int.}$} & \colhead{$F_{\rm 450\ \mu m}$} & \colhead{$F_{\rm 850\ \mu m}$} & \colhead{$\log(L_{\rm FIRST,Rakshit})$} & \colhead{$\sigma'_{\rm [OIII],Rakshit}$} & \colhead{$\log (M_*)$} \\
\colhead{ } & \colhead{ } & \colhead{(hh:mm:ss)} & \colhead{(dd:mm:ss)} & \colhead{ } & \colhead{ } & \colhead{(s)} & \colhead{(mJy)} & \colhead{(mJy)} & \colhead{$\mathrm{(erg\ s^{-1})}$} & \colhead{$\mathrm{(km\ s^{-1})}$} & \colhead{$\rm (M_\sun)$}\\
\rule{0pt}{0pt}  (1)&(2)&(3)&(4)&(5)&(6)&(7)&(8)&(9)&(10)&(11)&(12)
} 
\startdata 
1   & J0749 & 07:49:07 & $+$45:10:33.8 & 0.192 & 22B & 3760 & $<146$    & $28\pm2$ &$41.30 $& 299.36 & $10.63\pm0.32$ \\
2   & J0754 & 07:54:44 & $+$35:47:12.7 & 0.257 & 22B & 3752 & $<180$    & $<7$     &$41.10 $& 313.57 & $10.83\pm0.18$ \\
3   & J1119 & 11:19:02 & $+$31:51:22.5 & 0.262 & 22B & 3717 & $193\pm55$& $<7$     &$41.11 $& 532.79 & $10.89\pm0.15$ \\
4   & J1244 & 12:44:20 & $+$40:51:36.8 & 0.249 & 22B & 2383 & $<246$    & $<8$     &$42.08 $& 475.49 & $11.30\pm0.11$ \\
5   & J1449 & 14:49:21 & $+$42:21:01.2 & 0.179 & 22B & 3755 & $<275$    & $<9$     &$41.39 $& 588.05 & $10.80\pm0.09$ \\
6   & J1527 & 15:27:58 & $+$22:33:04.0 & 0.254 & 22B & 3758 & $<400$    & $28\pm3$ &$42.06 $& 308.29 & $10.61\pm0.23$ \\
7   & J1643 & 16:43:32 & $+$30:48:35.5 & 0.184 & 22B & 3760 & $<144$    & $<7$     &$41.02 $& 459.02 & $11.13\pm0.12$ \\
8   & J2351 & 23:51:56 & $-$01:09:13.3 & 0.174 & 23A & 3847 & $<673$    & $36\pm3$ &$41.79 $& 391.01 & $11.61\pm0.01$ \\
9   & J1105 & 11:05:39 & $+$02:02:57.3 & 0.106 & 23A & 3643 & $<653$    & $104\pm3$&$40.97 $& 326.34 & $11.05\pm0.11$ \\
10  & J1345 & 13:45:45 & $+$53:32:52.3 & 0.136 & 23A & 4265 & $161\pm48$& $160\pm2$&$40.79 $& 290.13 & $11.17\pm0.01$ \\
11  & J1557 & 15:57:07 & $+$45:07:00.1 & 0.234 & 23A & 3848 & $<181$    & $<7$     &$40.03 $& 341.11 & $10.32\pm0.12$ \\
12  & J0936 & 09:36:22 & $+$39:21:32.0 & 0.210 & 23A & 3863 & $<287$    & $<7$     &$40.79 $& 443.52 & $10.74\pm0.14$ \\
13  & J1153 & 11:53:24 & $+$58:31:38.5 & 0.202 & 23A & 4270 & $<158$    & $<6$     &$41.20 $& 317.05 & $11.12\pm0.11$ \\
14  & J1140 & 11:40:48 & $+$46:22:04.8 & 0.115 & 23A & 3862 & $187\pm53$& $11\pm2$ &$40.63 $& 596.84 & $11.12\pm0.16$ \\
15  & J0945 & 09:45:26 & $+$35:21:03.6 & 0.208 & 23A & 3653 & $<800$    & $<30$    &$41.49 $& 209.66 & $10.73\pm0.09$ \\
16  & J1532 & 15:32:29 & $+$04:53:58.3 & 0.218 & 23A & 3839 & $<132$    & $<7$     &$40.46 $& 317.53 & $10.71\pm0.24$ \\
17  & J1030 & 10:30:59 & $+$31:02:55.7 & 0.178 & 23A & 3649 & $<394$    & $61\pm3$ &$40.97 $& 247.50 & $11.14\pm0.21$ \\
18  & J2250 & 22:50:25 & $+$14:19:52.0 & 0.235 & 23A & 3651 & $<121$    & $61\pm2$ &$42.75 $& 275.02 & $10.99\pm0.11$ \\
19  & J1012 & 10:12:56 & $+$16:38:53.0 & 0.118 & 23A & 3651 & $230\pm70$& $<7$     &$40.40 $& 590.89 & $10.71\pm0.10$ \\
20  & J1005 & 10:05:08 & $+$34:14:24.1 & 0.162 & 23A & 3649 & $<362$    & $<8$     &$39.49 $& 342.05 & $9.33 \pm0.07$ \\
21  & J1027 & 10:27:36 & $+$28:59:17.6 & 0.257 & 23A & 1828 & $<355$    & $<11$    &$40.06 $& 287.39 & $10.33\pm0.10$ \\
22* & J1727 & 17:27:23 & $+$55:10:53.5 & 0.247 & ... & 10012& $<126$    & $36\pm1$ &$41.68 $& 510.30 & $11.73\pm0.11$ \\
23* & J1220 & 12:20:12 & $+$02:03:42.2 & 0.240 & 19A & 954  & $<303$    & $134\pm4$&$42.00 $& 402.17 & $11.79\pm0.10$ \\
24* & J1633 & 16:33:24 & $+$47:18:58.9 & 0.116 & 21B & 5420 & $<50 $    & $<5$     &$40.55 $& 360.33 & $10.57\pm0.14$ \\
25* & J1312 & 13:12:18 & $+$35:15:21.0 & 0.183 & ... & 4977 & $<27 $    & $5\pm1$  &$40.84 $& 529.90 & $11.28\pm0.16$ \\
26* & J1133 & 11:33:21 & $-$03:33:37.4 & 0.118 & 19A & 3787 & $<376$    & $<7$     &$38.82 $& 415.14 & $10.87\pm0.09$ \\
27* & J1202 & 12:02:27 & $-$01:29:15.2 & 0.150 & 20B & 3992 & $<290$    & $8\pm2$  &$40.09 $& 745.30 & $10.94\pm0.15$ \\
28* & J0159 & 01:59:50 & $+$00:23:40.9 & 0.163 & 20B & 1902 & $<220$    & $11\pm3$ &$40.44 $& 580.71 & $11.47\pm0.18$ \\
29* & J0207 & 02:07:13 & $-$01:12:23.1 & 0.172 & 19B & 5640 & $<435$    & $<7$     &$39.90 $& 439.06 & $11.06\pm0.10$ \\
30* & J0752 & 07:52:45 & $+$43:41:05.3 & 0.180 & 19A & 7557 & $<399$    & $<6$     &$40.22 $& 415.89 & $11.15\pm0.09$ \\
31* & J1356 & 13:56:18 & $-$02:31:01.4 & 0.134 & 19A & 5677 & $<323$    & $<6$     &$39.14 $& 326.79 & $10.78\pm0.09$ \\
32* & J0840 & 08:40:29 & $+$33:20:52.2 & 0.167 & 19A & 7568 & $<472$    & $<6$     &$39.51 $& 446.00 & $11.19\pm0.09$ \\
33* & J1025 & 10:25:31 & $+$51:40:34.8 & 0.045 & 20B & 2171 & $<39 $    & $<6$     &$37.62 $& 163.26 & $10.42\pm0.12$ \\
34* & J1215 & 12:15:49 & $+$54:42:24.0 & 0.150 & 20B & 3619 & $<234$    & $<7$     &$39.36 $& 457.84 & $11.01\pm0.22$ \\
35* & J0904 & 09:04:04 & $+$07:48:19.3 & 0.151 & 19A & 3781 & $<395$    & $<7$     &$39.34 $& 410.33 & $10.80\pm0.09$ \\
36* & J1614 & 16:14:13 & $+$26:04:16.1 & 0.131 & 19A & 2186 & $<196$    & $<8$     &$40.10 $& 238.70 & $10.86\pm0.26$ \\
37* & J1118 & 11:18:30 & $+$40:25:53.9 & 0.155 & 19A & 646  & $<282$    & $<14$    &$39.39 $& 479.77 & $11.35\pm0.23$ \\
38* & J0736 & 07:36:39 & $+$43:53:16.6 & 0.114 & 19A & 3784 & $<805$    & $<9$     &$39.35 $& 446.74 & $10.69\pm0.11$ \\
39* & J0747 & 07:47:04 & $+$48:38:33.3 & 0.220 & 21A & 3621 & $<76 $    & $<5$     &$39.48 $& 991.99 & $10.61\pm0.17$ \\
40* & J1422 & 14:22:30 & $+$29:52:24.2 & 0.113 & 19A & 3783 & $<418$    & $<7$     &$39.92 $& 537.15 & $10.90\pm0.10$ \\
41* & J1451 & 14:51:09 & $+$27:09:26.8 & 0.064 & 19A & 1888 & $<115$    & $<7$     &$38.75 $& 303.15 & $10.76\pm0.16$ \\
42* & J1509 & 15:09:14 & $+$17:57:10.0 & 0.171 & 19A & 1896 & $<431$    & $<9$     &$40.20 $& 382.18 & $11.10\pm0.10$ \\
\enddata
\tablecomments{Column (1): the target number, with asterisks indicating objects with archival SCUBA-2 data. The new objectes are arranged in the order of observations. Column (2): the abbreviated form of SDSS name. Columns (3)(4): the right ascension and declination in J2000 coordinates. Column (5): the redshift determined from SDSS spectra. Column (6): the observation semester. Column (7): the integration time of the JCMT observation. Columns (8)(9): the flux and its ${\rm 1 \sigma}$ uncertainty or upper limit at 450 ${\rm \mu m}$ and 850 ${\rm \mu m}$. Column (10): logarithm of the radio luminosity derived from the FIRST VLA 1.4 GHz survey from \cite{rakshit_census_2018}. Column (11): the outflow strength of [O III] from \cite{rakshit_census_2018}. Column (12): logarithm of the stellar mass of the host galaxy, which can be accessed from the \href{https://www.sdss3.org/dr10/spectro/galaxy_mpajhu.php}{MPA--JHU} database.}
\end{deluxetable*}

\subsection{SCUBA-2/JCMT Observations} \label{subsec:JCMTObservations}

The JCMT, located on Mauna Kea Mountain in Hawaii, features a 15-meter aperture and specializes in observations in the sub-mm band. One of its instruments, SCUBA-2 \citep{holland_scuba-2_2013}, is an imager capable of simultaneous operation at 450 $\rm\mu m$ and 850 $\rm\mu m$. During the 2022B and 2023A cycles, we used SCUBA-2 to observe seven and fourteen objects under the proposal IDs M22BP038 and M23AP023 (PI: Nguyen Nhat Kim Ngan), respectively. Details of these observations are documented in Table~\ref{tab:1ObsInfo}.

The effective FWHMs of the beams are $\rm 8.6~arcsec$ and $\rm 12.6~arcsec$ at $\rm 450~\mu m$ and $\rm 850~\mu m$ \citep{mairs_decade_2021}, respectively. Corresponding to physical sizes of $\rm 25.4~kpc$ and $\rm 37.2~kpc$ at the average redshift of 0.174 for our sample, the dust emission in the SCUBA-2 bands can be considered compact. So the DAISY mapping mode was employed in our observations. This mode maximizes exposure time at the center of the field and is specifically designed for compact or point-like sources. The majority of these observations were under band 3 weather conditions, which implies atmospheric opacities ranging from 0.08 to 0.12 at 225 GHz, though some were conducted under the clearer skies of band 2 or the slightly cloudier conditions of band 4. To avoid calibration loss during long observations, we allocated the observation time for each target into two minimal schedulable blocks, ensuring that none exceeded 40 minutes.

\subsection{Data Reduction} \label{subsec:DataReduction}

The raw data were processed using the \href{http://www.starlink.ac.uk/docs/sun258.htx/sun258ss40.html#xref_MAKEMAP}{makemap} command within the \href{http://starlink.eao.hawaii.edu/starlink/2021ADownload}{Starlink 2021A} software suite \citep{currie_starlink_2014}, which is included in the \href{http://www.starlink.ac.uk/docs/sun258.htx/sun258.html#xref_}{SMURF} package \citep{chapin_scuba-2_2013}. Considering that all our targets are compact sources, we employed the \href{https://raw.githubusercontent.com/Starlink/starlink/master/applications/smurf/examples/dimmconfig_blank_field.lis}{dimmconfig\_blank\_field.lis} configuration, optimized for detecting extremely faint point sources in blank fields. Subsequently, images were calibrated using the default Flux Conversion Factor, and combined images of the same targets were merged into a single image following the \href{http://www.starlink.ac.uk/docs/sun265.htx/sun265.html#xref_}{PICARD} \href{http://www.starlink.ac.uk/docs/sun265.htx/sun265ss6.html#Q1-13-32}{CALIBRATE\_SCUBA2\_DATA} and the \href{http://www.starlink.ac.uk/docs/sun265.htx/sun265ss15.html#Q1-22-59}{MOSAIC\_JCMT\_IMAGES} recipes.

To maximize the visibility of point sources, a matched filter was applied to the combined images using the \href{http://www.starlink.ac.uk/docs/sun265.htx/sun265.html#xref_}{PICARD} \href{http://www.starlink.ac.uk/docs/sun265.htx/sun265ss24.html#Q1-31-86}{SCUBA2\_MATCHED\_FILTER} recipe, with default settings of 20 arcsec and 30 arcsec at ${\rm 450\ \mu m}$ and ${\rm 850\ \mu m}$, respectively. The images were then cropped to 180 arcsec using the \href{http://www.starlink.ac.uk/docs/sun265.htx/sun265.html#xref_}{PICARD} \href{http://www.starlink.ac.uk/docs/sun265.htx/sun265ss11.html#Q1-18-47}{CROP\_SCUBA2\_IMAGES} recipe to avoid false signal detection at the edges.

SNR maps were generated from the cropped images using the \href{http://www.starlink.ac.uk/docs/sun95.htx/sun95.html}{KAPPA} \href{http://www.starlink.ac.uk/docs/sun95.htx/sun95ss109.html#Q1-136-553}{MAKESNR} command, which creates a new map by dividing the DATA component by the square root of its VARIANCE component. We then contoured SNR levels of 3 and 2. A target was considered significantly detected if its SDSS position fell within a contour of 3. The flux was determined by the peak value at the clump, and uncertainty was estimated from the corresponding value on the error image produced by \href{http://www.starlink.ac.uk/docs/sun258.htx/sun258ss40.html#xref_MAKEMAP}{MAKEMAP}. If no clump larger than 3 but a clump larger than 2 was found at the SDSS position, we re-applied \href{http://www.starlink.ac.uk/docs/sun265.htx/sun265ss24.html#Q1-31-86}{SCUBA2\_MATCHED\_FILTER} and \href{http://www.starlink.ac.uk/docs/sun265.htx/sun265ss11.html#Q1-18-47}{CROP\_SCUBA2\_IMAGES} with a filter size set to three times the effective FWHM derived from a two-component fit of the beams, i.e., 25.8 arcsec and 37.8 arcsec at ${\rm 450\ \mu m}$ and ${\rm850\ \mu m}$, respectively, to enhance the SNR. If a new clump with SNR $>$ 3 was detected at the SDSS position, it was also considered as detected. If there is no detection yet, we assigned three times the maximum error value of the cropped image as the flux upper limit. Table~\ref{tab:1ObsInfo} lists the detected fluxes and upper limits for non-detections of our 42 targets at ${\rm 450\ \mu m}$ and ${\rm 850\ \mu m}$.

\subsection{Multiband Data} \label{subsec:31MultiwavelengthData}

We gathered flux and uncertainty estimations from the VizieR database \citep{ochsenbein_vizier_2000}. Our photometric data cover FUV to FIR bands: GALEX (FUV and NUV bands, \citealt{bianchi_revised_2017}), SDSS (optical $\it ugriz$ bands, \citealt{ahumada_16th_2020}), ALLWISE (Near-infrared $\mathrm{JHK_s}$ bands and MIR W1-4 bands, \citealt{cutri_allwise_2014}), IRAS (MIR and FIR bands, \citealt{helou_infrared_1988, moshir_iras_1990, wang_imperial_2009, saunders_pscz_2000}), Herschel (FIR bands, \citealt{viero_herschel_2014, valiante_herschel_2016, shangguan_gas_2018, brown_spectral_2019}), and JCMT ($\mathrm{450\ \mu m }$ and $\mathrm{850\ \mu m}$ bands of SCUBA-2). The majority of our objects have photometric data from GALEX, SDSS, and ALLWISE. However, IRAS data were available for only 10 objects, and observations by Herschel were limited to 8 objects. Given the vital role of FIR data in estimating SFRs, the JCMT sub-mm data are particularly important.

\subsection{Radio Data} \label{subsec:radiodata}

AGNs are often considered as point sources from X-ray to IR bands due to the limited resolution of most instruments. In contrast, a significant fraction of radio-loud AGNs shows extended structures in radio images with arcsec resolution. Catalogs of these extended sources typically list fitted Gaussian components, complicating the matching process between high-resolution and low-resolution catalogs, which means a point source in a low-resolution catalog could correspond to multiple components in a high-resolution catalog. Accurately estimating jet emission at SCUBA-2 bands needs radio data that cover a broad frequency range. However, the varied resolutions across catalogs present significant challenges. To simplify this process, data from SPECFIND V3.0 \citep{stein_specfind_2021} were utilized.

SPECFIND is a hierarchical code that organizes matched flux densities into various classifications such as siblings, parents, and children, based on their spatial proximity and flux densities at the same and different frequencies. Then it fits their spectral slopes when there are enough data to guarantee consistency and uniformity across the observations. SPECFIND V3.0 matched data from 204 radio catalogs, which span a broad frequency from 16.7 MHz to 31 GHz, covering most of the sky.

When the radio data cover few frequencies, SPECFIND does not fit the data. Therefore, to ensure completeness, we collected more data from four additional catalogs: VLASS (2--4~GHz, with a resolution of 2.5 arcsec, \citealt{gordon_quick_2021}), FIRST (1.4~GHz, 6.8 arcsec, \citealt{helfand_last_2015}), and GMRT (147~MHz, 25 arcsec, \citealt{intema_gmrt_2017,de_gasperin_radio_2018}).

Additionally, we matched our sample with Planck (30--857~GHz, \citealt{planck_collaboration_planck_2018}). The lower spatial resolution (7.3 arcminutes at 143GHz and 32.3 arcminutes at 30GHz) of the Planck satellite limited its use to a reference of blazars. Within a radius of 7.3 arcsec around our sample, 6 objects are matched.

\subsection{Comparison Sample of Type-II AGNs}

To complement our sample of type-I AGNs and to provide a comparison sample for several of our analyses, we use the sample of type-II AGNs from \citet{kim_determining_2022}. Their sample consists of 39 type-II AGNs with $z < 0.2$, selected to cover a broad distribution of ionized-gas outflow strengths based on the [O III] $\lambda5007$ line kinematics from \citet{2016ApJ...817..108W}. Using new JCMT sub-mm observations, combined with archival multiband data, \citet{kim_determining_2022} performed detailed SED fitting to estimate dust-based SFRs. Because their measurements of SFRs, outflow properties, and AGN luminosities are similar with our following methods, this type-II sample can be considered as a comparison sample. We use it to compare properties of host galaxies of type-I and type-II AGNs and to study whether trends related to jet activity and star formation are similar across different AGN types.

\section{Analyses} \label{sec:3Methodology}

\begin{figure*}[ht]
\includegraphics[width=1.0\linewidth]{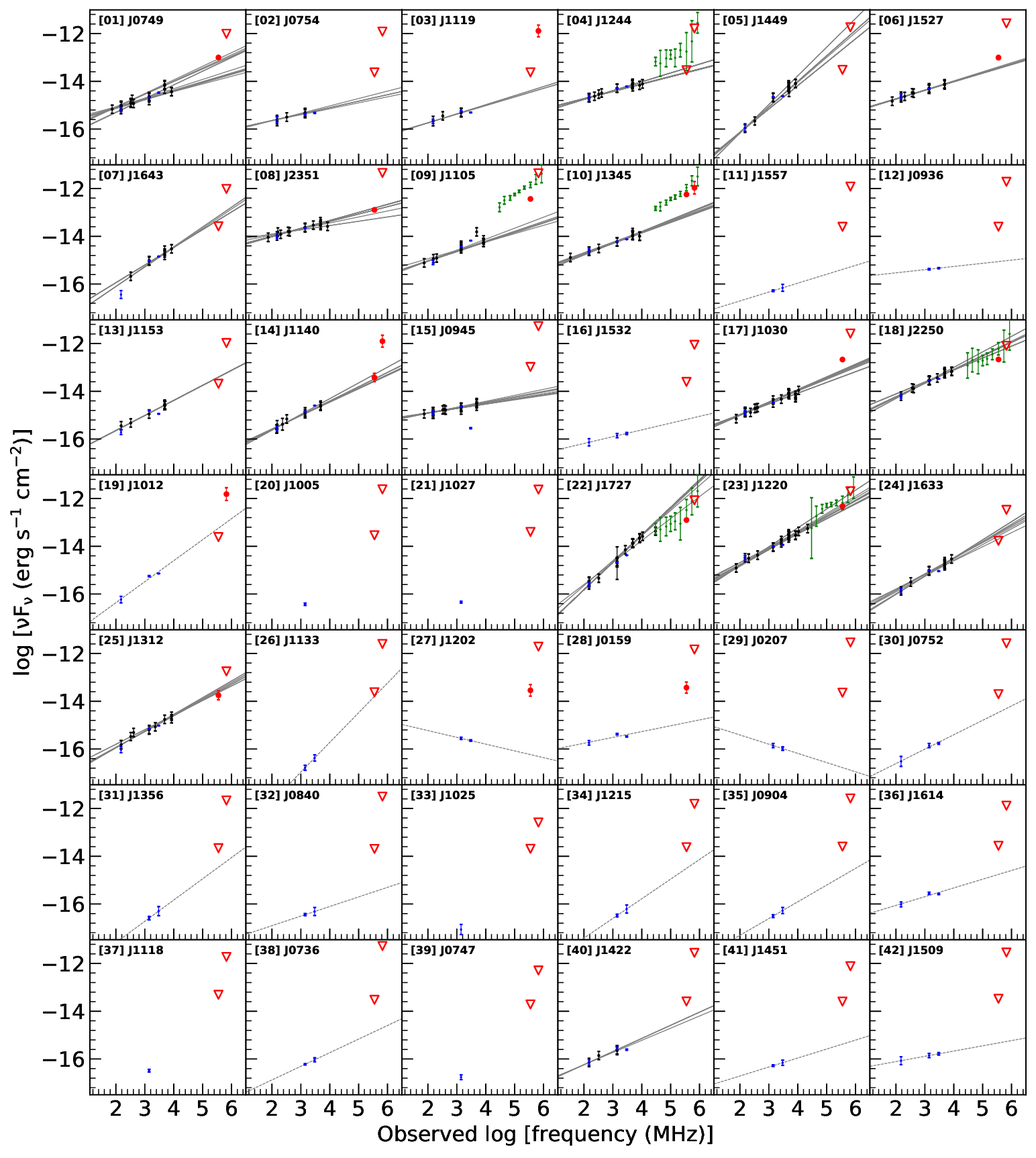}
\centering
\caption{Radio and SCUBA-2 data and best-fit models of each target. Each panel includes radio data, with blue and green dots representing data from SPECFIND and Planck and black dots indicating data from other catalogs. Additionally, SCUBA-2 observations are denoted by red dots for detected data and downward red triangles for upper limits. The gray solid lines represent the best-fitting models derived from the SPECFIND dataset, while gray dashed lines depict the best-fits using linear models for the other 4 additional catalogs.
\label{fig:3Radio}}
\end{figure*}

In this section, we assess the contribution of jets in the SCUBA-2 bands, calibrate the radio luminosity at 1.4 GHz, perform SED fittings, and fit the SDSS spectra of our sample. These analyses enable us to calculate SFRs using various methods and estimate the properties of AGNs and their host galaxies.

\subsection{Jet Emission} \label{subsec:32RemoveFSRQs}

As suggested by \cite{kennicutt_star_1998}, one relatively reliable indicator of SFR is the FIR luminosity between rest-frame 8 ${\rm \mu m }$ and 1000 ${\rm \mu m }$, $L_{\rm 8-1000 \mu m}$. This is based on the assumption that dust heated by AGNs is typically hotter than that by stars \citep{xu_agns_2020}, and the radiation from the hot dust is mainly in the MIR band, dropping steeply in the FIR band. Therefore, the FIR luminosity can be adopted to estimate SFRs. This method is probably feasible for radio-quiet AGNs. However, \cite{rojas-ruiz_impact_2021} suggested that the jets of luminous radio-loud AGNs might also contribute to sub-mm radiation. For some of radio-loud AGNs of our sample, synchrotron emission may extend into the sub-mm band, potentially contaminating the SCUBA-2 fluxes and leading to an overestimation of the dust emission and SFR. Therefore, it is necessary to model the radio–submm continuum in order to (i) obtain accurate estimates of the jet luminosity and (ii) identify and correct cases where jet emission contributes significantly to the observed sub-mm data. These steps ensure that the sub-mm fluxes used in SED fitting truly reflect dust emission from star formation.

The radio and sub-mm data are presented in Figure~\ref{fig:3Radio}. In Figure~\ref{fig:3Radio}, we show SPECFIND data as black points and data from VLASS, FIRST, GMRT, and Planck as blue points. SPECFIND fits the spectrum through multiple iterations. For the uniformity of the radio data and the reliability of the fitting results, data will be excluded if they deviate significantly from the current spectral fit or if their frequency is too close to other data points. To ensure that each data point is considered, these excluded points can be combined with other data to attempt to form a new linear fitting. Therefore, for one object, SPECFIND may provide multiple fitting results. This process continues until all data are either described by a linear fitting result or it is determined through iteration that no suitable result can describe them. The gray solid lines in Figure~\ref{fig:3Radio} show the fitting results of SPECFIND. We use the mean values of the flux densities estimated by all fitted spectra at the SCUBA-2 bands to predict the contribution of the jet in the sub-mm band.

When there are few data or even for a subset of the data there is not a stable linear fitting result, spectral fitting results from SPECFIND were unavailable. Therefore we used the four additional catalogs in Subsction~\ref{subsec:radiodata} to fit spectra and estimate the jet emission at the SCUBA-2 bands as a supplement to SPECFIND. The gray dashed lines in Figure~\ref{fig:3Radio} show our results.

We calculated logarithmic ratios between the observed SCUBA-2 data (including upper limits) and the predicted flux of the jet at 450 ${\rm \mu m}$ and 850 ${\rm \mu m}$ bands. To avoid contamination from jet emission in sub-mm data, we replaced values with ratios less than 0.5 dex (a linear ratio of 3.16) as upper limits, which might indicate a significant jet contribution and hard to distinguish dust emission at the sub-mm band. Then for objects with ratios larger than 0.5 dex, we compared the predicted SCUBA-2 data with the uncertainties of the detected data. If the uncertainty was larger, the influence of jet emission was negligible. Conversely, if it was smaller, we subtracted the predicted values from the observed data to remove the contribution of the jets.

In Figure~\ref{fig:3Radio}, the Planck data of 6 matched objects are shown as green dots. Due to the low resolution of Planck data, it is only considered when its values closely match the SCUBA-2 data within the margin of error. ID04, being upper limits, requires no further analysis of jet emission. ID18, ID22, and ID23 are replaced by upper limits based on the above criteria. Planck data of ID09 are disregarded due to a significant discrepancy between Planck and SCUBA-2 data, suggesting a crossmatching error. For ID10, the consistency of Planck and SCUBA-2 data within the error margins identifies it as a blazar, and it is then replaced as upper limits. In Table~\ref{tab:2Radio}, $F^{\rm corr}_{\rm 450}$ and $F^{\rm corr}_{\rm 850}$ present the corrected SCUBA-2 data.

\subsection{1.4~GHz Emission}\label{subsec:1.4GHz}

In \cite{rakshit_census_2018}, only the ``CORE" component of the radio emission from FIRST was considered, and the K-correction was not taken into account. However, to accurately evaluate the jet power, a reliable estimate of the total intrinsic 1.4 GHz luminosity is required, because $L_{\rm 1.4GHz}$ is a standard tracer of jet activity and is widely used in comparisons with previous works. This differs from Section~\ref{subsec:32RemoveFSRQs}, where the radio-submm spectral fitting is used to estimate the jet contribution at SCUBA-2 wavelengths rather than to measure the total jet luminosity.

Although the FIRST catalog has deeper data and higher spatial resolution than the NRAO VLA sky survey (NVSS, \citealt{condon_nrao_1998}), it is less sensitive to extended radio components \citep{white_signals_2007}. And sometimes faint extended radio emission is completely missing in the FIRST catalog \citep{2003NewAR..47..593B}. \cite{zhu__l_2020} also found that for some FIRST resolved extended radio sources, the flux of FIRST is significantly lower than that of NVSS by 30--50 per cent. Therefore, we crossmatch our sample with NVSS catalog by a radius of 10 arcsec and prioritize using NVSS flux density to avoid underestimating the total radio emission, provided that it is not contaminated by very nearby background radio sources, and when NVSS data are not available, we use FIRST data as a supplement. 

We calculate the radio emission at 1.4GHz and apply the K-correction using the following formula:
\[L_{\rm 1.4GHz} = \nu_R F_{\rm int} \times \frac{1}{(1+z)^{1+\alpha}}  \times (4 \pi D_L^2), \]
where $\nu_R$ is the frequency of $\rm 1.4~GHz$, $F_{\rm int}$ is the integrated radio flux of the NVSS or FIRST survey, and $D_L$ is the luminosity distance at the redshift of $z$. The K-correction is needed to convert the observed flux to its rest-frame luminosity, and the spectral index $\alpha$ ensures the proper shape of the synchrotron spectrum.

For the spectral index of $\alpha$, we use the linear fitting results obtained in Subsection~\ref{subsec:32RemoveFSRQs}. When it is not available, we use the typical 1.4--3 GHz spectral index of $-0.71$, as provided by the comparison of the VLASS at $\nu \sim 3$ GHz and the FIRST in \cite{gordon_quick_2021}. This approach provides a more accurate estimate of the intrinsic 1.4 GHz luminosity than the values reported by \cite{rakshit_census_2018}, which do not include extended emission or a K-correction.

By comparing $L_{\rm 1.4GHz}$ and $L_{\rm FIRST,Rakshit}$, we find that for the four clearly extended objects of IDs 08, 10, 17, 23, the difference is about 0.43 dex, indicating that the catalog of \cite{rakshit_census_2018} severely underestimates the emission of extended radio objects. $L_{\rm 1.4GHz}$ is shown in Table~\ref{tab:2Radio}.

\subsection{SED Fitting} \label{subsec:33SEDFitting}

\begin{table}[t!] 
\centering
\caption{List of the Input Parameters of CIGALE} \label{tab:2Models}
\begin{tabularx}{8.5cm}{lc}
\hline\hline
{Parameters} & {Input Values}\\
\hline
\multicolumn{2}{c}{SFH: sfhdelayed}\\
\hline
{tau\_main} & {1,500,1000,2000,4000,8000}\\
{age\_main} & {10000}\\
{tau\_burst}& {20000}\\
{age\_burst}& {5,10,50,100,500,1000}\\
{f\_burst}  & {0.0,0.01,0.05,0.1,0.5}\\
\hline
\multicolumn{2}{c}{SSP: bc03}\\
\hline
{imf}        & {Chabrier}\\
{metallicity}& {0.02}\\
\hline
\multicolumn{2}{c}{Dust attenuation: dustatt\_calzleit}\\
\hline
{E\_BVs\_young}      & {0.05,0.1,0.15,0.2,0.4,0.6,0.8,1.0}\\
{E\_BVs\_old\_factor}& {0.44,1.0}\\
{powerlaw\_slope}    & {$-0.3$,0.0,0.3}\\
\hline
\multicolumn{2}{c}{Dust Emission: dl2014}\\
\hline
{umin}  & {0.1,0.5,1.0,10.0}\\
{alpha} & {2.0,2.1,2.3,2.5,3.0}\\
\hline
\multicolumn{2}{c}{AGN Emission: skirtor2016}\\
\hline
{t}      & {3,7,11}\\
{i}      & {30,70}\\
{fracAGN}& {(1) 0.01,0.05,0.1,0.2,0.3,0.4,0.5,0.8,0.9}\\
{}       & {(2) 0.01,0.05,0.1,0.2,0.3,0.4,0.5}\\
{EBV}    & {0.03,0.05,0.1,0.3}\\
\hline
\end{tabularx}
\end{table}

To reliably derive dust luminosities and dust-based SFRs, and to decompose the contributions of AGN-heated dust and star-formation-heated dust, we perform SED fitting using the Code Investigating GAlaxy Emission (CIGALE, \citealt{noll_analysis_2009, boquien_cigale_2019, yang_fitting_2022}). CIGALE provides a suite of options including star-formation history (SFH) models, along with dust attenuation and emission, and AGN modules. CIGALE calculates a grid of SED models, and provides Bayesian-like estimations for model parameters.

The parameters of CIGALE models are listed in Table~\ref{tab:2Models}. In this study, we focus on dust luminosity instead of nebulae, so we have set the parameters of the nebular emission model to default values: the ionization parameter (${\rm log}\ U$) is fixed at ${\rm -2.0}$, both the fraction of Lyman continuum photons escaping the galaxy ($f_{\rm esc}$) and absorbed by dust ($f_{\rm dust}$) are set to zero, and the line width is established at 300 ${\rm km/s}$. 

For the emission of the host galaxy, we employed the Simple Stellar Population (SSP) model by \cite{bruzual_stellar_2003}, the IMF from \cite{chabrier_galactic_2003}, and delayed SFH models with an optional exponential burst. The metallicity was set to the solar value of 0.02. CIGALE models adhere to the principle of energy conservation where emission absorbed in the optical and UV bands by dust is re-emitted in the IR spectrum. Because this study focuses on SFR and dust luminosity, parameters related to dust attenuation and emission are of most importance. The dust attenuation curve follows the formulations by \cite{calzetti_dust_2000, leitherer_global_2002} and with the color excess of the continuum light of the young stellar population (E\_BVs\_young) presented in Table~\ref{tab:2Models}. The reduction factor for the ${\rm E(B-V)}$ of the older stellar population relative to the younger one, denoted as E\_BVs\_old\_factor, was set to values of 0.44 and 1.0. For dust emission, we selected the model by \cite{draine_andromedas_2014}, specifying a mass fraction of 0.47 for Polycyclic Aromatic Hydrocarbons and a fraction illuminated from Umin to Umax at 0.1. The input values of the minimum radiation field (umin) and the slope (alpha) of the function of the radiation field ${\rm (d}U{/\rm d}M \propto U^\alpha)$ are provided in Table~\ref{tab:2Models}. AGN contributions were simulated using the clumpy torus models developed by \cite{stalevski_3d_2012, stalevski_dust_2016}.

We find that for IDs 05, 12, 28, 36, 38, and 41, the best-fit CIGALE models show that the contribution of AGNs and host galaxies to the FIR radiation can be comparable or even FIR emission is dominated by AGNs. This contradicts the physical picture that FIR radiation is primarily from the cold dust of host galaxies. Therefore, for these sources, we have constrained $\rm fracAGN \leq 0.5$, as shown in $\rm fracAGN(2)$ in Table~\ref{tab:2Models}.

\begin{figure*}[ht]
\includegraphics[width=1.0\linewidth]{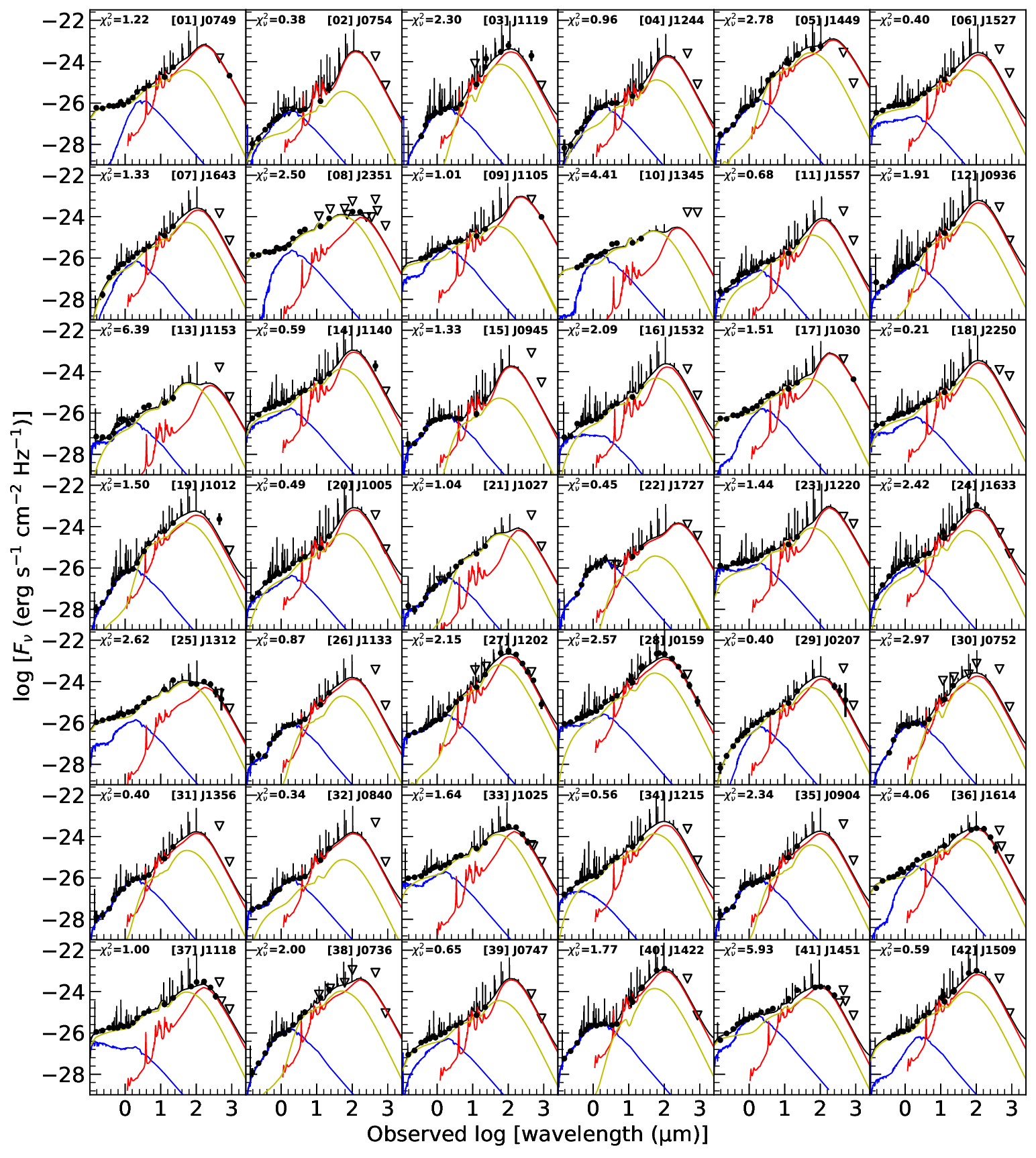}
\caption{The best-fit CIGALE models for our sample. Blue, yellow, red, and black curves denote the attenuated galaxy, AGN, dust components, and the total models, respectively. Observed data points are marked with black-filled dots and upper limits with open triangles. Reduced ${\chi^2}$ values are displayed in the upper-left corner of each panel.
\label{fig:4SED}}
\end{figure*}

Our best-fit SED models are presented in Figure~\ref{fig:4SED}. The detected photometric data are denoted by black dots, while the upper limits are indicated by black open triangles. The black curve represents the model that best fits the data, with the contributions from the attenuated stellar, AGN, and dust components in blue, yellow, and red curves, respectively. The reduced $\chi^2$ values, mostly below 3 and with a median of 1.38, are indicated in the upper left corner of each panel, suggesting the general agreement between our models and data.

CIGALE provides Bayesian-like posterior probability distributions for model parameters, such as SFRs based on SFH ($\rm SFR_{0Myr}$ and $\rm SFR_{100Myr}$), and we also used the Bayesian value of the dust luminosity to calculate the SFRs following the formula:
\[
{\rm log\ SFR\ (M_\sun\ yr^{-1})} = {\rm log}\ L_{\rm Dust}\ {\rm(erg\ s^{-1})}-43.591.
\]
where $L_{\rm Dust}$ comes from the fitting results of CIGALE and represents the re-emission of radiation absorbed in star-forming regions. \cite{kennicutt_star_1998} first introduced this formula, and \cite{kim_determining_2022} also used the dust luminosity obtained from CIGALE to estimate SFRs. The ${\rm SFR_{Dust, SED}}$ results are shown in Table~\ref{tab:2Radio}.

\begin{deluxetable*}{cccccccccccccc}[ht!]
\tabletypesize{\scriptsize}
\setlength{\tabcolsep}{1.0mm}{}
\tablecaption{SFRs, Radio, and $M_{\rm BH}$ Information \label{tab:2Radio}}
\tablehead{
\colhead{ID} & \colhead{Short name} & \colhead{Full SDSS name} & \colhead{$\sigma'_{\rm [OIII]}$}& \colhead{$F^{\rm corr}_{\rm 450}$} & \colhead{$F^{\rm corr}_{\rm 850}$} & \colhead{Model} & \colhead{$\alpha$} & \colhead{$\log(L_{\rm 1.4GHz})$} & \colhead{${\rm \log (SFR_{Dust,SED})}$}& \colhead{$E(B-V)$}& \colhead{$ \log (M_{\rm BH})$}\\
\colhead{ } &\colhead{ } &\colhead{ } & \colhead{${\rm (km\ s^{-1})}$} & \colhead{${\rm (mJy)}$} & \colhead{${\rm (mJy)}$} &\colhead{ } & \colhead{ }& \colhead{${\rm (erg\ s^{-1})}$} & \colhead{${\rm (M_\odot\ yr^{-1})}$}& \colhead{ }& \colhead{${\rm (M_\odot)}$}\\
\rule{0pt}{0pt}  (1)&(2)&(3)&(4)&(5)&(6)&(7)&(8)&(9)&(10)&(11)&(12)
} 
\startdata 
1   & J0749 & J074906.50+451033.8 & $287\pm9$   & $146$	     & $21\pm2^\dag$ & 1 & $ -0.57 $ & 41.30 & $ 1.81\pm0.21 $  & $ 0.63 $ & $ 8.34 $   \\
2   & J0754 & J075444.08+354712.7 & $337\pm5$   & $181$	     & $7$           & 1 & $ -0.73 $ & 40.92 & $ 1.53\pm0.29 $  & $ 0.47 $ & $ 7.51^* $ \\
3   & J1119 & J111902.26+315122.5 & $592\pm17$  & $193\pm55$ & $7$             & 1 & $ -0.63 $ & 40.99 & $ 2.18\pm0.06 $  & $ 0.00 $ & $ 7.52^* $ \\
4   & J1244 & J124419.96+405136.8 & $473\pm12$  & $246$	     & $8$           & 1 & $ -0.67 $ & 41.95 & $ 1.19\pm0.31 $  & $ 0.21 $ & $ 7.45^* $ \\
5   & J1449 & J144920.71+422101.2 & $663\pm7$   & $275$	     & $9$           & 1 & $  0.06 $ & 41.22 & $ 2.12\pm0.03 $  & $ 1.01 $ & $ 7.38^* $ \\
6   & J1527 & J152757.67+223304.0 & $309\pm9$   & $400$	     & $28$          & 1 & $ -0.64 $ & 41.93 & $ 1.88\pm0.31 $  & $ 0.01 $ & $ 8.21 $   \\
7   & J1643 & J164331.91+304835.5 & $484\pm17$  & $145$	     & $7$           & 1 & $ -0.22 $ & 40.90 & $ 1.54\pm0.24 $  & $ 1.05 $ & $ 8.88^* $ \\
8   & J2351 & J235156.12-010913.3 & $558\pm27$  & $673$	     & $36$          & 1 & $ -0.70 $ & 42.26 & $ 0.84\pm0.02 $  & $ 0.58 $ & $ 8.40 $   \\
9   & J1105 & J110538.99+020257.3 & $635\pm12$  & $653$	     & $97\pm3^\dag$ & 1 & $ -0.60 $ & 40.99 & $ 1.15\pm0.07 $  & $ 0.29 $ & $ 7.93 $   \\
10  & J1345 & J134545.36+533252.3 & $592\pm110$ & $162$	     & $160$         & 1 & $ -0.54 $ & 41.44 & $ 0.31\pm0.49 $  & $ 0.00 $ & $ 8.03 $   \\
11  & J1557 & J155707.14+450700.1 & $348\pm5$   & $181$	     & $7$           & 2 & $ -0.63 $ & 39.79 & $ 1.02\pm0.32 $  & $ 0.24 $ & $ 7.70^* $ \\
12  & J0936 & J093621.51+392132.0 & $517\pm3$   & $287$	     & $7$           & 2 & $ -0.87 $ & 40.66 & $ 2.01\pm0.06 $  & $ 0.93 $ & $ 8.30^* $ \\
13  & J1153 & J115323.96+583138.5 & $298\pm5$   & $158$	     & $6$           & 1 & $ -0.37 $ & 40.88 & $ 0.25\pm0.23 $  & $ 0.44 $ & $ 8.00 $   \\
14  & J1140 & J114047.89+462204.8 & $760\pm16$  & $188\pm53$ & $11$            & 1 & $ -0.40 $ & 40.64 & $ 1.27\pm0.36 $  & $ 0.57 $ & $ 7.96 $   \\
15  & J0945 & J094525.90+352103.6 & $202\pm5$   & $800$	     & $30$          & 1 & $ -0.79 $ & 41.42 & $ 1.25\pm0.20 $  & $ 0.09 $ & $ 7.53^* $ \\
16  & J1532 & J153228.79+045358.3 & $340\pm5$   & $132$	     & $7$           & 2 & $ -0.72 $ & 40.34 & $ 1.45\pm0.07 $  & $ 0.35 $ & $ 8.36^* $ \\
17  & J1030 & J103059.09+310255.7 & $309\pm8$   & $394$	     & $43\pm3^\dag$ & 1 & $ -0.49 $ & 41.42 & $ 1.69\pm0.11 $  & $ 0.39 $ & $ 8.30 $   \\
18  & J2250 & J225025.34+141952.0 & $261\pm2$   & $121$	     & $61$          & 1 & $ -0.44 $ & 42.61 & $ 1.63\pm0.32 $  & $ 0.21 $ & $ 7.93 $   \\
19  & J1012 & J101256.03+163853.0 & $598\pm5$   & $230\pm70$ & $7$             & 2 & $ -0.12 $ & 40.28 & $ 1.40\pm0.21 $  & $ 0.19 $ & $ 6.92^* $ \\
20  & J1005 & J100508.08+341424.1 & $354\pm4$   & $362$	     & $8$           & 0 & $ -0.71 $ & 39.42 & $ 1.64\pm0.19 $  & $ 0.73 $ & $ 8.10^* $ \\
21  & J1027 & J102736.45+285917.6 & $297\pm8$   & $355$	     & $11$          & 0 & $ -0.71 $ & 39.94 & $ 0.84\pm0.30 $  & $ 0.00 $ & $ 6.92^* $ \\
22  & J1727 & J172723.47+551053.5 & $574\pm63$  & $126$	     & $36$          & 1 & $ 0.13 $  & 41.46 & $ 1.31\pm0.27 $  & $ 0.10 $ & $ 7.12^* $ \\
23  & J1220 & J122011.88+020342.2 & $437\pm19$  & $303$	     & $134$         & 1 & $ -0.32 $ & 42.15 & $ 1.93\pm0.14 $  & $ 0.84 $ & $ 8.53 $   \\
24  & J1633 & J163323.58+471858.9 & $305\pm9$   & $50$	     & $5$           & 1 & $ -0.30 $ & 40.50 & $ 1.62\pm0.02 $  & $ 0.97 $ & $ 7.11 $   \\
25  & J1312 & J131217.75+351521.0 & $644\pm15$  & $26$	     & $5$           & 1 & $ -0.34 $ & 40.75 & $ 0.83\pm0.02 $  & $ 1.92 $ & $ 8.47 $   \\
26  & J1133 & J113320.56-033337.5 & $442\pm10$  & $376$	     & $7$           & 2 & $  0.24 $ & 38.71 & $ 0.96\pm0.14 $  & $ 0.21 $ & $ 6.83^* $ \\
27  & J1202 & J120226.75-012915.2 & $824\pm10$  & $290$	     & $8\pm2$       & 2 & $ -1.28 $ & 40.11 & $ 2.11\pm0.02 $  & $ 1.18 $ & $ 6.61^* $ \\
28  & J0159 & J015950.25+002340.8 & $709\pm10$  & $220$	     & $11\pm3$      & 2 & $ -0.76 $ & 40.41 & $ 2.30\pm0.03 $  & \nodata  & $ 8.00 $   \\
29  & J0207 & J020712.81-011221.8 & $383\pm11$  & $435$	     & $7$           & 2 & $ -1.39 $ & 39.84 & $ 1.28\pm0.07 $  & $ 0.72 $ & $ 7.39^* $ \\
30  & J0752 & J075244.63+434105.3 & $415\pm3$   & $399$	     & $6$           & 2 & $ -0.40 $ & 40.16 & $ 1.57\pm0.07 $  & $ 0.77 $ & $ 6.95^* $ \\
31  & J1356 & J135617.79-023101.5 & $323\pm4$   & $323$	     & $6$           & 2 & $ -0.11 $ & 39.04 & $ 1.18\pm0.10 $  & $ 0.26 $ & $ 6.74^* $ \\
32  & J0840 & J084028.61+332052.2 & $451\pm8$   & $472$	     & $6$           & 2 & $ -0.60 $ & 39.41 & $ 1.44\pm0.13 $  & $ 0.29 $ & $ 6.71^* $ \\
33  & J1025 & J102531.28+514034.8 & $187\pm13$  & $39$	     & $6$           & 0 & $ -0.71 $ & 37.60 & $-0.10\pm0.03 $  & $ 2.62 $ & $ 7.06 $   \\
34  & J1215 & J121549.43+544223.9 & $411\pm10$  & $234$	     & $7$           & 2 & $ -0.18 $ & 39.25 & $ 1.57\pm0.05 $  & $ 1.13 $ & $ 6.80 $   \\
35  & J0904 & J090403.72+074819.3 & $429\pm5$   & $395$	     & $7$           & 2 & $ -0.30 $ & 39.24 & $ 1.22\pm0.11 $  & $ 0.00 $ & $ 6.47^* $ \\
36  & J1614 & J161413.20+260416.3 & $236\pm3$   & $196$	     & $8$           & 2 & $ -0.63 $ & 40.04 & $ 1.31\pm0.02 $  & $ 0.37 $ & $ 7.83 $   \\
37  & J1118 & J111830.28+402554.0 & $785\pm16$  & $282$	     & $14$          & 0 & $ -0.71 $ & 39.31 & $ 0.99\pm0.02 $  & $ 2.96 $ & $ 7.68 $   \\
38  & J0736 & J073638.86+435316.6 & $479\pm3$   & $805$	     & $9$           & 2 & $ -0.43 $ & 39.12 & $ 1.29\pm0.02 $  & $ 0.13 $ & $ 6.81^* $ \\
39  & J0747 & J074704.43+483833.3 & $675\pm22$  & $76$	     & $5$           & 0 & $ -0.71 $ & 39.36 & $ 1.46\pm0.30 $  & $ 1.08 $ & $ 7.88 $   \\
40  & J1422 & J142230.34+295224.2 & $599\pm8$   & $418$	     & $7$           & 1 & $ -0.46 $ & 39.88 & $ 1.76\pm0.03 $  & $ 0.55 $ & $ 7.49^* $ \\
41  & J1451 & J145108.76+270926.9 & $274\pm5$   & $115$	     & $7$           & 2 & $ -0.62 $ & 38.61 & $ 0.55\pm0.02 $  & $ 1.57 $ & $ 7.25 $   \\
42  & J1509 & J150913.79+175710.0 & $361\pm9$   & $431$	     & $9$           & 2 & $ -0.78 $ & 40.11 & $ 1.99\pm0.03 $  & $ 0.47 $ & $ 7.72 $   \\
\enddata
\tablecomments{Column (1): target number. Column (2): shortening of SDSS name. Column (3): full SDSS name. Column (4): the outflow strength of [O III] fitted by BADASS. Columns (5)(6): corrected flux and uncertainty at 450 $\rm \mu m$ and 850 $\rm \mu m$. Note that some flux data are replaced as upper limits because of substantial contamination of jet emission, and the dagger means that jet emission is more significant than uncertainty so it is subtracted from flux. Column (7): the model used when fitting radio data. 0 indicates that there is not enough data to fit, 1 represents the results from SPECFIND, and 2 represents the linear fitting results from the four additional catalogs. Column (8): the spectral index obtained from fitting radio data. When the model is 0, we use the $\alpha$ of $-0.71$ from \cite{gordon_quick_2021}. Column (9): logarithm of the radio luminosity at $\rm 1.4\ GHz$ derived from Subsection~\ref{subsec:1.4GHz}. Column (10): logarithm of SFRs based on dust luminosity of SED fitting. Column (11): the dust extinction derived from Balmer decrement of the narrow components of $\rm H\alpha$ and $\rm H\beta$ emission lines. Column (12): logarithm of SMBH mass by single-epoch method, with an asterisk indicating the absence of broad ${\rm H\beta}$ components, the details of which are discussed in Appendix~\ref{A2:MBH}.}
\end{deluxetable*}

\subsection{Spectral Analysis} \label{subsec:34Spectra}

To measure SMBH masses using single-epoch methods and estimate the dust in the narrow-line region (NLR), we used the Bayesian AGN Decomposition Analysis for SDSS Spectra (BADASS, \citealt{sexton_bayesian_2020}) to fit the SDSS spectra of our sample. Additionally, we fit the spectra of type-II AGNs of \cite{kim_determining_2022}. They selected a sample of 39 type-II AGNs with a broad distribution of outflows from \cite{2016ApJ...817..108W}. After obtaining sub-mm data by JCMT and fitting the SEDs by CIGALE, they estimated the SFRs in the same way as we do. Therefore, we use their data as an important supplement to our sample. BADASS can simultaneously fit a variety of spectral components, which include the power-law continuum, the Balmer and iron pseudo continuum, narrow and broad components of the emission lines, and the emission of the host galaxies. The fitting process employs Markov Chain Monte Carlo (MCMC) methods to ensure robust uncertainty estimation.

\begin{figure*}[ht]
\includegraphics[width=1.0\linewidth]{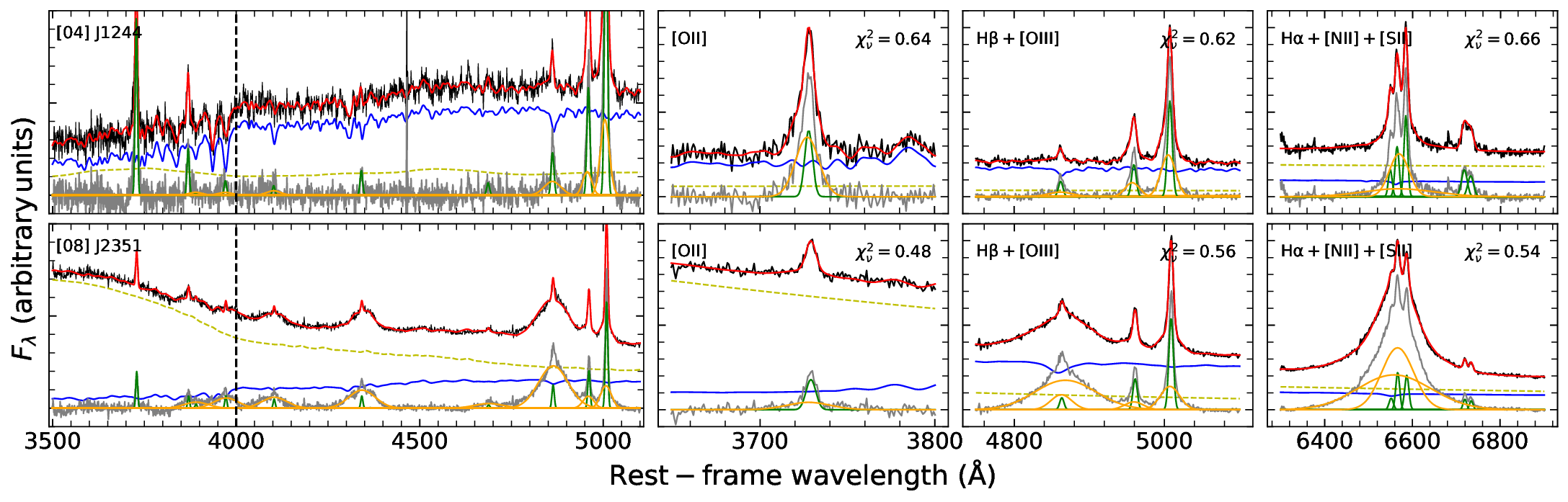}
\caption{Typical SDSS spectra of host-dominated (top panel) and AGN-dominated (bottom panel) types of our sample. Across the panels, from left to right, each column presents the continuum (3500--5100${\rm~\AA}$), [O II] ${\rm \lambda 3727}$ (3650--3800${\rm~\AA}$), ${\rm H\beta +[O\ III]}$ (4750--5100${\rm~\AA}$), and ${\rm H\alpha + [N\ II]+[S\ II]}$ (6300--6900${\rm~\AA}$). The observed spectra, best-fit results, and residuals are depicted by black, red, and gray curves. Furthermore, the contributions from the AGN continuum, the host galaxy, and narrow and broad emission lines, are shown using yellow dashed, blue, green, and orange solid curves, respectively. The vertical black dashed lines mark the position at ${\rm 4000~\AA}$, indicating the location of the break feature, which can be utilized for estimating the SFR. 
\label{fig:5Spectra}}
\end{figure*}

To avoid the extensive amount of time required for global spectral fitting with BADASS, we focus on four wavelength regions and fit them separately:

1. The continuum (3500--5100${\rm~\AA}$). Considering that the features of the AGN and host galaxy continuum are predominantly found at shorter wavelengths, we employed a simple model, including the host galaxy continuum, power law, iron continuum, and Balmer continuum, to fit the continuum spectrum. For the Balmer lines and the ${\rm [O\ III]\ \lambda 4959}$ and ${\rm [O\ III]\ \lambda 5007}$ emission lines, we used two Gaussian profiles, while other emission lines were fitted with a single Gaussian profile.

2. [O II] ${\rm \lambda 3727}$ (3650--3800${\rm~\AA}$). \cite{zhuang_recalibration_2019} suggested that after correcting for metallicity and dust attenuation, the [O II] ${\rm \lambda 3727}$ emission line can serve as an estimator for the SFR. Similar to the continuum fitting in range (1), we employed a model consisting of the host galaxy, power-law, UV iron continuum, and Balmer continuum to fit the continuum emission, together with two Gaussian components to model the [O II] ${\rm \lambda 3727}$ line within the ${\rm 3650-3800\AA}$ spectra.

3. ${\rm H\beta +[O\ III]}$ (4750--5100${\rm~\AA}$). We modeled the continuum using a three-component model (host galaxy, power-law, and iron continuum). We used up to three Gaussian models (one narrow and two broad components) to accurately represent broad ${\rm H\beta}$, while two Gaussians (one narrow and one broad) were used for ${\rm [O\ III]}$. The FWHM of the entire broad ${\rm H\beta}$ profile is considered as indicative of the kinematics of the broad line region (BLR) clouds, which can be used to estimate the mass of the SMBH via the single-epoch method, the specifics of which are detailed in Appendix~\ref{A2:MBH}.

4. ${\rm H\alpha + [N\ II]+[S\ II]}$ (6300--6900${\rm~\AA}$). We modeled the continuum using the same model as in range (3) and employed three Gaussians (one narrow and two broad) to model the ${\rm H\alpha}$ emission and a single narrow Gaussian for other lines (${\rm [N\ II]}$ and ${\rm [S\ II]}$). The luminosity of the ${\rm H\alpha}$ line is used to estimate the SFR. In cases where broad ${\rm H\beta}$ components are absent, the FWHMs of the broad profiles of ${\rm H\alpha}$ are used as upper limits for the kinematics of the BLR. Given the strong ionized gas outflows of our sample, it is possible that ${\rm [N\ II]}$ might exhibit outflow components as well, contributing to the broad components around ${\rm H\alpha + [N\ II]}$. However, the close proximity of wavelengths makes it difficult to distinguish the broad ${\rm H\alpha}$ and ${\rm [N\ II]}$ outflows. Consequently, the luminosity and FWHM of ${\rm H\alpha}$ might be overestimated. Further discussions are in Appendix~\ref{A2:MBH}.

For type-II AGNs as discussed by \cite{kim_determining_2022}, we fitted the spectra across the same four ranges mentioned above. We used a combination of host galaxy emission and a power-law model to fit the continuum. To model the ${\rm [O\ II]}$, ${\rm [O\ III]}$, and ${\rm H\alpha}$ lines, two Gaussians were employed, while a single Gaussian was used for ${\rm H\beta}$, ${\rm [N\ II]}$, and ${\rm [S\ II]}$.

\cite{osterbrock_astrophysics_1989} suggested that under dust-free Case B conditions (electron temperatures $T_e = {\rm 10^4\ K}$ and electron densities $n_e = {\rm10^2–10^4\ cm^{-3}}$), the intrinsic line ratio of $\rm H\alpha$ to $\rm H\beta$ for AGNs is 3.1, which is known as the Balmer decrement. By using the observed line ratios of the narrow components of $\rm H\alpha$ and $\rm H\beta$, and assuming the Milky Way extinction curve of \cite{cardelli_relationship_1989} with $R_V = A_V / E(B-V) = 3.1$, we can estimate the amount of dust extinction $E(B-V)$ in the NLR, and correct the luminosity of emission lines. If the observed line ratio is smaller than 0, the corresponding $E(B-V)$ is set to 0. The results of $E(B-V)$ are shown in Table~\ref{tab:2Radio}.

The fraction of the host component ($f_{\rm Host}$) at $\rm 4000~\AA$, derived from fitting the continuum, exhibits a bimodal distribution with a saddle point of approximately 0.4. Therefore, in Figure~\ref{fig:5Spectra}, we present the spectra and fitting results for two representative objects with $f_{\rm Host}$ values of 0.78 and 0.22, respectively. The upper panel displays J1244, where the continuum is predominantly dominated by the emission of the host galaxy, whereas the lower panel features J2351, characterized by an AGN-dominated continuum. Sequentially from left to right, the four columns present the fitting results of the continuum, [O II] ${\rm \lambda 3727}$, ${\rm H\beta +[O\ III]}$, and ${\rm H\alpha + [N\ II]+[S\ II]}$, respectively. The observed spectra, best-fit models, and residuals are represented by black, red, and grey curves. The contributions of the AGN continuum, host galaxy, and narrow and broad components of emission lines, are depicted with yellow dashed, blue, green, and orange solid curves, respectively. Vertical black dashed lines denote the ${\rm 4000 \AA}$, where the break feature is useful for SFR estimation.

\section{Results and Discussions} \label{sec:4Results}

SFR estimation has considerable uncertainties ($\sim$ 0.7 dex, as mentioned by \citealt{merloni_cosmic_2010}), due to the systematics of different measurement methods such as UV, optical, and IR measurements. We calculated and compared SFRs using five methods: the ${\rm 4000~\AA}$ break (${\rm SFR_{Dn4000}}$), dust luminosity (${\rm SFR_{Dust,SED}}$), the SFH (${\rm SFR_{0Myr}}$ and ${\rm SFR_{100Myr}}$), ${\rm H\alpha}$ emission (${\rm SFR_{H\alpha}}$, \citealt{kennicutt_star_1998}), and ${\rm [O\ II]}$ emission lines (${\rm SFR_{[O II]}}$, \citealt{zhuang_recalibration_2019}). All optical measurements (Dn4000, H$\alpha$, and [O II]) are derived from the AGN–host decomposed spectra obtained with BADASS (see Figure~\ref{fig:5Spectra}), ensuring that continuum and line fluxes used for SFR estimates are corrected for AGN contamination.

We found that the ${\rm SFR_{Dust,SED}}$ is generally larger than the SFRs obtained by other methods. However, due to the potential contamination of ${\rm SFR_{Dn4000}}$, ${\rm SFR_{H\alpha}}$, and ${\rm SFR_{[OII]}}$ in AGN-dominated spectra, and since ${\rm SFR_{0Myr}}$ and ${\rm SFR_{100Myr}}$ might largely depend on the choice of CIGALE models, we weigh that ${\rm SFR_{Dust, SED}}$, which is based on multi-band data and dust luminosity, can more accurately and robustly reflect the star formation for our sample. Because minor AGN contamination may persist even after decomposition, the dust-based SFR provides a more robust measure for our analysis. In Paper I, we compared ${\rm SFR_{Dust,SED}}$ and ${\rm SFR_{Dn4000}}$ as an example, in which it can be found that ${\rm SFR_{Dust,SED}}$ is statistically larger than ${\rm SFR_{Dn4000}}$.

We describe the measurements of the properties of SMBHs in Appendix~\ref{A2:MBH}. Not all the targets of our sample show broad ${\rm H\beta}$ components. When these are missing, we use ${\rm H\alpha}$ to study the kinematics of the BLR. We discuss how the mix of broad ${\rm H\alpha}$ components and [N II] outflows affects our measurements. Also, we describe how to estimate the mass or the upper limit mass of the SMBH using broad components of ${\rm H\beta}$ and ${\rm H\alpha}$ and the relationship between SMBH mass and stellar mass. The SMBH masses for our study are listed in Table~\ref{tab:2Radio}, with an asterisk marking where broad ${\rm H\beta}$ components are missing.

\subsection{${SFR_{Dust,SED}}$ and Luminosity} \label{subsec:44SFR_IR_FIR}

\begin{figure*}[t!]
\includegraphics[width=1.0\linewidth]{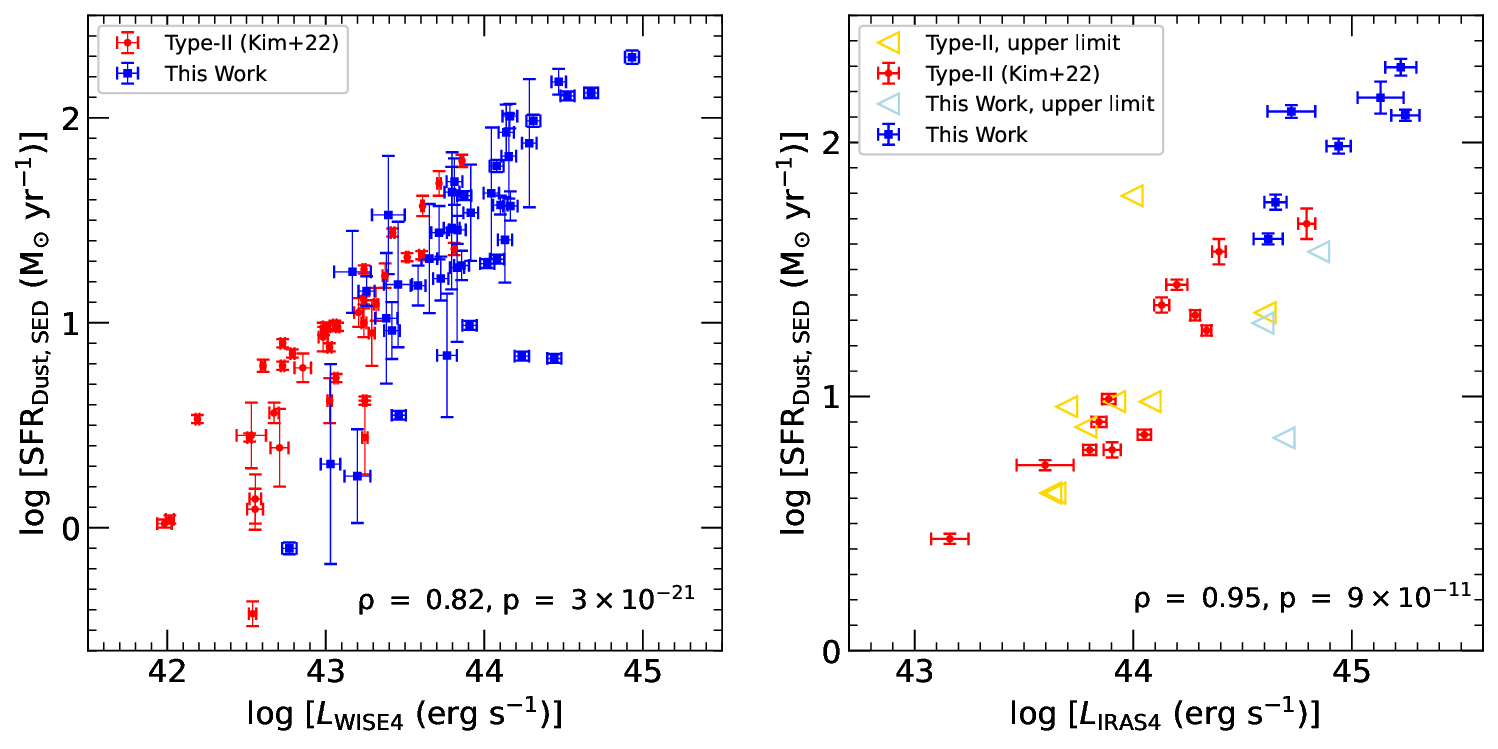}
\caption{Comparison of the SFRs based on the dust luminosity with the luminosities at the WISE4 band (left panel) and the IRAS4 band (right panel) for type-I (represented by blue squares, our sample) and type-II (indicated by red circles, from \citealt{kim_determining_2022}) AGNs. Red circles and blue squares denote targets with detection at corresponding IR bands, while yellow and light blue triangles represent targets for which only upper limits are available. The Spearman's rank correlation coefficients and $p$-values of detected objects are shown in the lower-right corners.
\label{fig:9FIR}}
\end{figure*}

\begin{figure*}[t!]
\includegraphics[width=1.0\linewidth]{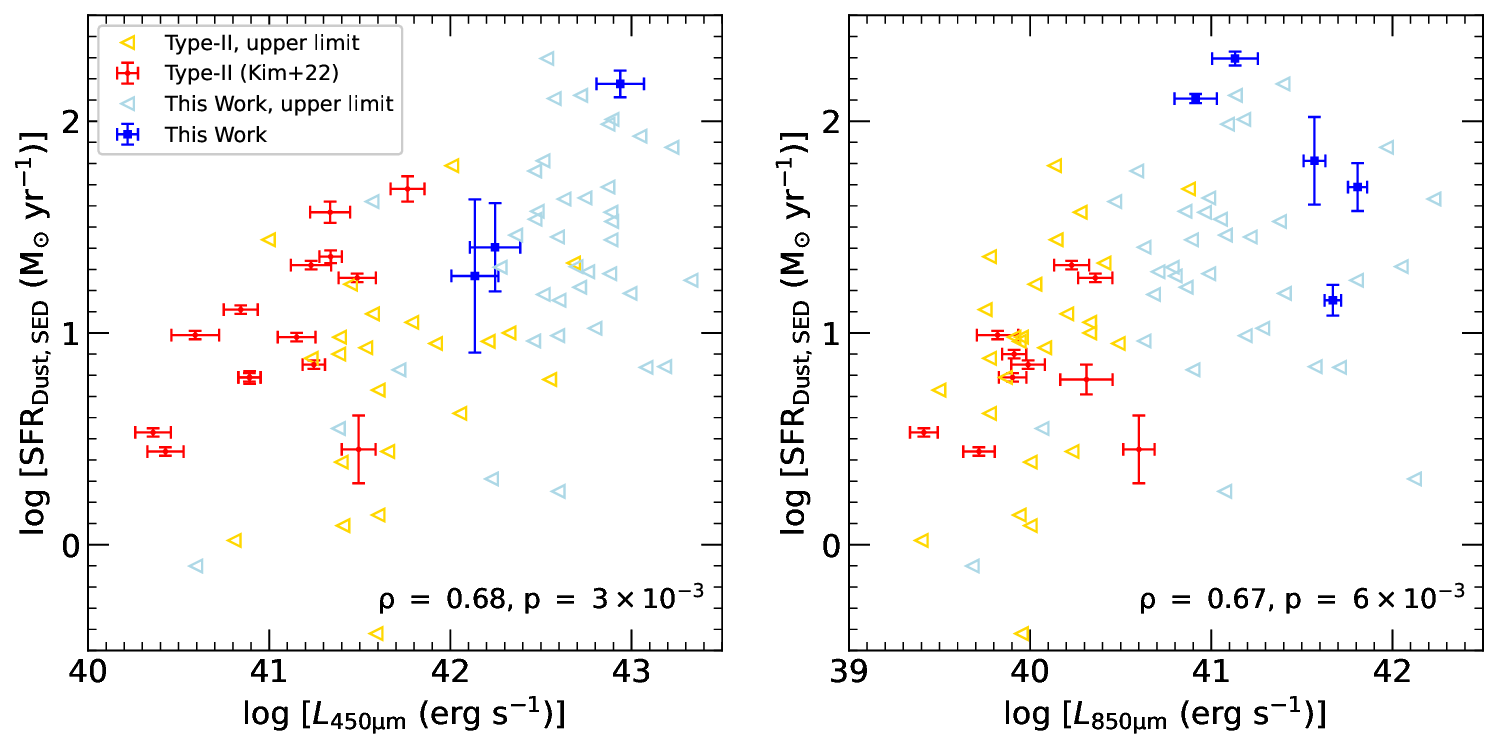}
\caption{Comparison of the SFRs based on dust luminosity with the luminosities at ${\rm 450\ \mu m}$ (left panel) and ${\rm 850\ \mu m}$ (right panel) for type-I (represented by blue squares, our sample) and type-II (indicated by red circles) AGNs. The Spearman's rank correlation coefficients and $p$-values are also from only detected objects.
\label{fig:10SCUBA2}}
\end{figure*}

In Figure~\ref{fig:9FIR}, we show the relations between ${\rm SFR_{Dust, SED}}$ and the luminosities of two IR bands. Red-filled circles and blue-filled squares identify type-II and type-I AGNs detected in the corresponding FIR bands, whereas yellow and light blue open triangles signify targets for which only upper limits exist. For the IRAS4 band, with an effective wavelength (${\lambda_{\rm ref}}$) of ${\rm 100\ \mu m}$, a significant linear correlation is evident, which is also quantified by the Spearman's rank correlation coefficient. Likewise, for the WISE4 band with ${\lambda_{\rm ref}}$ of ${\rm 22\ \mu m}$, a positive correlation is also found.

It is generally accepted that for objects with a strong AGN component, the luminosity at the MIR band is predominantly dominated by radiation from the torus. However, as shown in Figure~\ref{fig:4SED}, for some objects (such as ID02, ID22, ID26, etc.), the AGN components displayed in the SEDs are notably weak, which makes the luminosity at the WISE4 band still primarily from dust components, and results in the observed positive correlation, albeit with large scatter.

In Figure~\ref{fig:10SCUBA2}, we compare ${\rm SFR_{Dust,SED}}$ with the luminosities at two SCUBA-2 bands. Despite significant scatter, luminosities at both 450 and 850 ${\rm \mu m}$ exhibit a consistent linear relationship with the SFRs. The Spearman's rank correlation coefficients and the corresponding $p$-values are displayed in the lower-right corners of their respective panels. The $p$-values indicate that the correlation coefficients are statistically significant. Although there is a correlation between SCUBA-2 luminosities and SFRs, it is with larger scatter compared to the luminosity at 100 ${\rm \mu m}$ from IRAS4. This discrepancy may be attributed to the fact that most dust re-emission is concentrated around ${\rm 100~\mu m}$. The dust models at about ${\rm 850~\mu m}$ are so diverse, which may lead to poor representation of dust emission at this band. Therefore, we do not recommend relying solely on the luminosity at ${\rm 850~\mu m}$ for accurate estimation of SFRs in nearby AGNs.

\subsection{Jetted AGNs} \label{subsec:42jetted}

\begin{figure}[t!]
\includegraphics[width=1.0\linewidth]{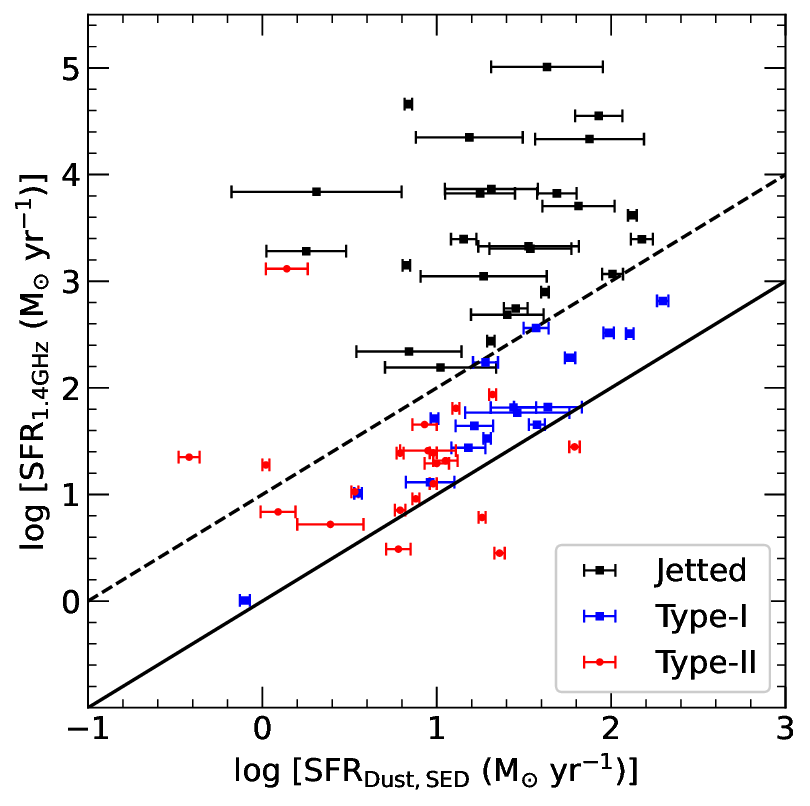}
\caption{Comparison of the SFRs based on dust luminosity obtained from CIGALE with the SFRs predicted from the radio power for type-I (represented by blue and black squares, our sample) and type-II (indicated by red dots, from \citealt{kim_determining_2022}) AGNs. The black squares are classified as jetted AGNs because their radio emission should be dominated by jets. The blue squares are other type-I AGNs of our sample. Their radio emission might come from the star formation. The black solid line and the black dashed line represent $\rm SFR_{1.4GHz} > SFR_{Dust, SED}$ and $\rm SFR_{1.4GHz} > 10\times SFR_{Dust, SED}$, respectively.
\label{fig:SFRjet}}
\end{figure}

The influence of the jets, specifically their correlation with SFRs, is the focus of this work. Therefore, distinguishing whether the radio emission in our sample originates from the jets or from star formation is very important. \cite{yun_radio_2001} suggested that SFRs can be estimated using the radio power at 1.4 GHz as follows:
\[\rm SFR_{1.4GHz}\ (M_{\odot}\ yr^{-1})= 5.9\times10^{-22}P_{1.4GHz}(W\ Hz^{-1}),\]
where $\rm P_{1.4GHz} = L_{1.4GHz}/(1.4GHz)$ represents the radio power. We multiply the $\rm SFR_{radio}$ by a factor of 0.6 to convert the IMFs from \cite{salpeter_luminosity_1955} to \cite{chabrier_galactic_2003}. \cite{bonzini_star_2015} showed that for star-forming galaxies and radio-quiet AGNs, the SFRs estimated from radio and FIR have a very good linear relationship. However, for radio-loud AGNs, the $\rm SFR_{1.4GHz}$ are significantly larger, likely due to the contribution of the jets to the radio power.

In Figure~\ref{fig:SFRjet}, we compare the $\rm SFR_{1.4GHz}$ with the $\rm SFR_{Dust,SED}$ for our type-I sample (represented as blue and black squares) and type-II AGNs of \cite{kim_determining_2022} (indicated as red dots). All type-I AGNs are located above the black solid line, which means $\rm SFR_{1.4GHz} > SFR_{Dust, SED}$, similar to the radio-loud AGNs in \cite{bonzini_star_2015}. Conservatively, we consider sources with $\rm SFR_{1.4GHz} > 10\times SFR_{Dust, SED}$ (represented by the black dashed line) as jetted AGNs, meaning that the radio emission primarily comes from the jet. This allows us to study the relationship between jet emission and star formation. For type-II AGNs, we match with NVSS using a matching radius of 5 arcsec and assume $\alpha = -0.71$ to estimate the radio power. Most red dots are located near the black solid line, suggesting that their radio emission is likely contributed mainly by star formation. Due to the scarcity of radio detection for type-II AGNs, we ignore the radio emission for these sources.

\subsection{Star Formation Main Sequence} \label{subsec:51Mainsequence}

\begin{figure}[t!]
\includegraphics[width=1.0\linewidth]{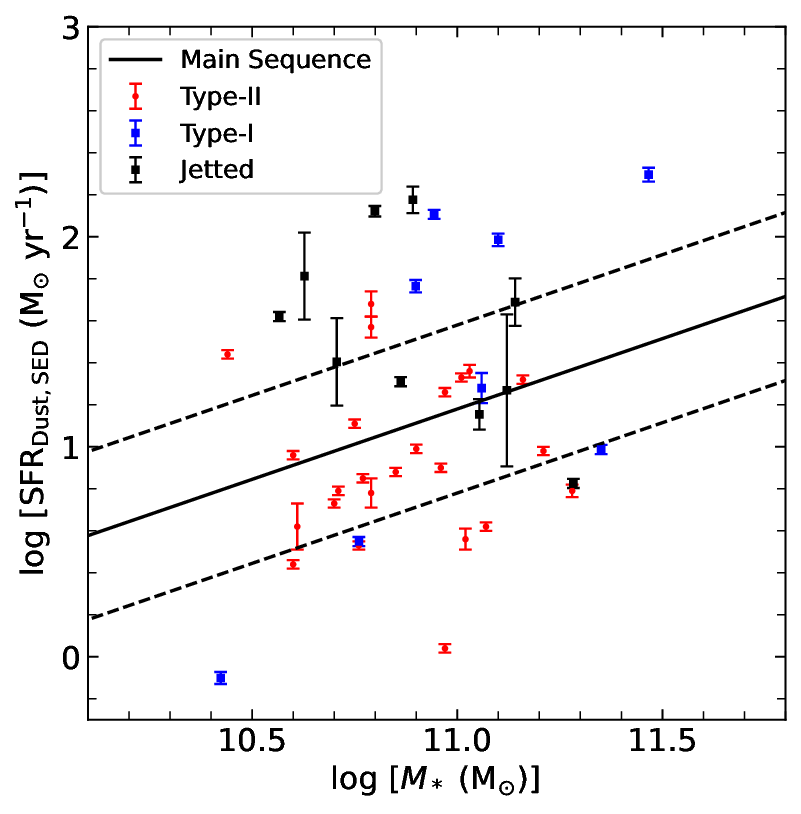}
\caption{The SFRs as a function of the stellar mass for type-II (red circles) and type-I AGNs. Type-I AGNs are divided into jetted and other objects and shown as black and blue squares. We only consider  AGNs with FIR or sub-mm data to ensure the reliability of the SFRs. The black solid and dashed lines represent the main sequence and its ${\rm 1 \sigma}$ dispersion, respectively, which is from \cite{noeske_star_2007}. 
\label{fig11:MS}}
\end{figure}

\cite{noeske_star_2007} presented the main sequence of star formation. Accordingly, in Figure~\ref{fig11:MS}, we compare SFRs and stellar masses, with the main sequence shown by a black solid line and its ${\rm 1\sigma}$ dispersion by two dashed lines, described by the equation:
\[{\rm log\ SFR_{MS}=(0.67\pm0.08)\times log\ M_*-(6.19\pm0.78)}.\]
The red circles represent type-II AGNs from \cite{kim_determining_2022}, and the black squares indicate jetted type-I AGNs which we select in the above subsection, and blue squares represent other type-I AGNs of our sample.

For type-I and type-II AGNs, we only consider objects with FIR or sub-mm detections to ensure that SFRs are reliable. Besides, through spectral observations using the Double Spectrograph instrument on the Palomar 200-inch telescope, we found that the spectra of ID~08 are quite unusual. As a type-I AGN, it has very red continuum radiation, which cannot be explained by the radiation of its host galaxy. Therefore, to avoid contamination, we excluded this source from the following analyses. In AGNSTRONG IV, we will present our study about the kinematics and properties of this object.

In Figure~\ref{fig11:MS}, we find that 15 out of 24 type-II AGNs are located below the main sequence. For the blue squares, 5 out of 8 are above the black solid line. We think there may be a selection effect. Because we selected objects with FIRST detection, and if the radio emission for the blue squares mostly comes from star formation, objects with higher SFRs are more likely to be selected in our sample. Besides, we found that jetted type-I AGNs (8 out of 10) predominantly exceed the main sequence, which might suggest a potential positive correlation between the jet and star formation, and we will discuss the possible mechanism in the following subsection.

\subsection{Is There More Dust?} \label{subsec:feedbackormoregas}

Two main mechanisms can explain that most objects with high relative radio luminosities are above the main sequence. One is that the positive feedback of the jets promotes star formation, and another is that objects with more gas and dust have more fuel supply, which results in higher SFRs and stronger jets. To determine which one is more dominant for our sample, we compared the dust extinction $E(B-V)$ obtained from the Balmer decrement in Subsection~\ref{subsec:34Spectra} and the distance of the objects from the star formation main sequence, i.e., $\rm log\ SFR_{Dust, SED} - log\ SFR_{MS}$.

\begin{figure}[t!]
\includegraphics[width=1.0\linewidth]{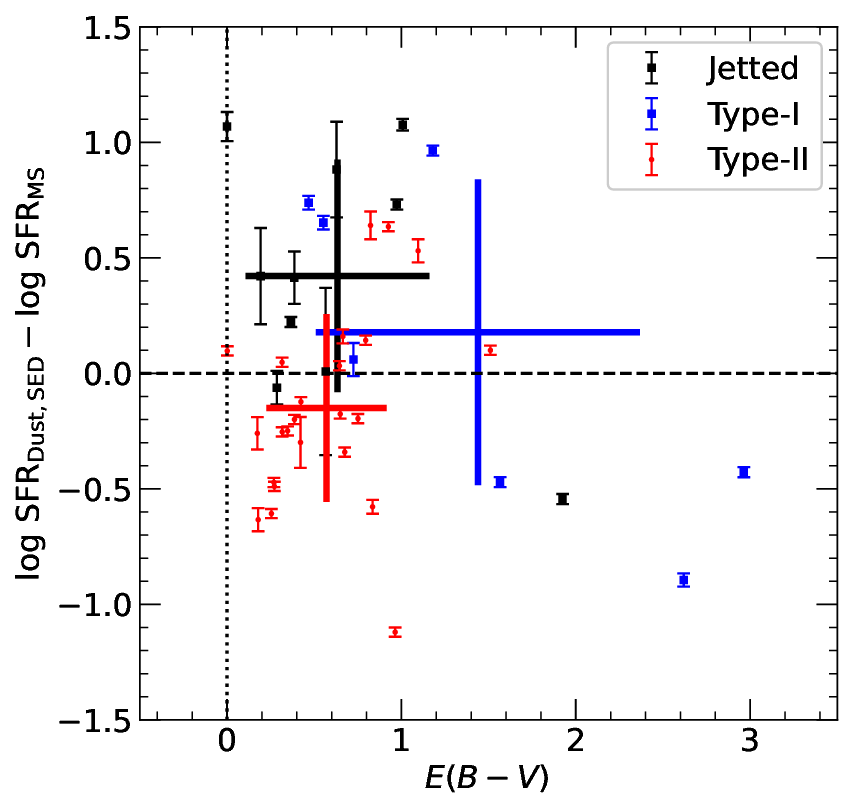}
\caption{Comparison of the dust extinction $E(B-V)$ with the SFR enhancement for type-I (represented by black and blue squares) and type-II (indicated by red dots, from \citealt{kim_determining_2022}) AGNs, with large pluses indicating the mean values and standard deviations for each group. The black dashed horizontal line and the black dotted vertical line represent $E(B-V)$ and SFR enhancement of 0, respectively.
\label{fig:Dustjet}}
\end{figure}

In Figure~\ref{fig:Dustjet}, the blue and black squares have the same meaning as in Figure~\ref{fig11:MS}. Assuming that the gas-to-dust ratio is roughly the same in our sample, we can use the dust extinction $E(B-V)$ to represent the amount of gas and dust in the NLR. If the positive correlation between radio luminosity and SFRs in Figure~\ref{fig11:MS} is due to the gas and dust supply of these sources, then the farther a source is above the star formation main sequence, i.e., the larger $\rm log\ SFR_{Dust, SED} - log\ SFR_{MS}$, the larger $E(B-V)$ should be, and vice versa. 

While a strong trend is not seen in our sample, a mild increase of $E(B-V)$ with distance from the main sequence may be present for our sample, and the small sample size limits our ability to determine whether a clear correlation exists. Therefore, the current data do not allow us to exclude the possibility that enhanced gas and dust supply contributes to their enhanced SFRs. Positive feedback remains a plausible interpretation, but we emphasize that it should be regarded as a tentative explanation rather than a definitive one.

For other type-I AGNs, there seems to be a trend that the larger $E(B-V)$, the smaller $\rm log\ SFR_{Dust, SED} - log\ SFR_{MS}$, but this sample is smaller and the relationship is more diffuse, possibly due to random phenomena.

\subsection{Correlation of sSFRs with [O III] Outflows} \label{subsec:52SFROutflow}

\begin{figure*}[ht!]
\includegraphics[width=1.0\linewidth]{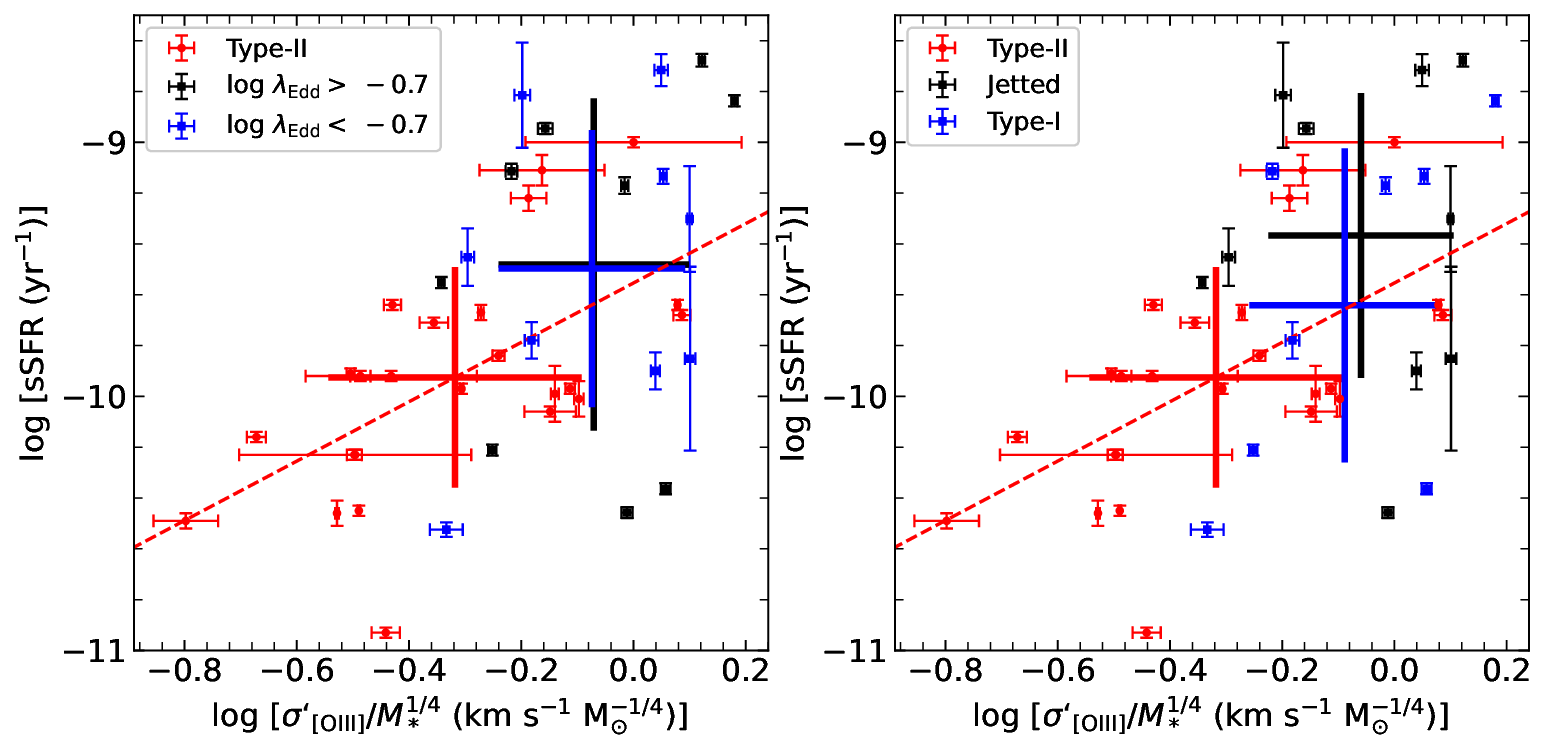}
\caption{The sSFRs as a function of the [O III] outflow strength for type-II (red circles) and type-I (blue and black squares) AGNs, with large pluses indicating the mean values and standard deviations for each group. The red dashed line is the linear relation for type-II AGNs. We only consider objects with FIR or sub-mm detections for type-I AGNs. In the left panel, we divide our type-I sample into two sub-samples with high and low Eddington ratios, and in the right panel, we consider jetted and other type-I AGNs respectively.
\label{fig12:Outflow}}
\end{figure*}

\cite{woo_delayed_2017} presented a positive correlation between the sSFRs and the [O III] outflow strength of type-II AGNs in spiral galaxies. In Figure~\ref{fig12:Outflow}, we show the same trend for type-II AGNs (represented by red dots) from \cite{kim_determining_2022} alongside our type-I AGNs (denoted by blue and black squares). The linear fitting results for the type-II AGNs are indicated by a red dashed line. Our sample is further divided based on high and low Eddington ratios in the left panel, and in the right panel, we represent jetted AGNs as black squares and indicate other type-I AGNs as blue squares, with large pluses indicating the mean values and standard deviations for each group.

The Eddington ratios of most type-II AGNs are below 0.1 (22 out of 24 objects in total have $\rm log\ \lambda_{Edd} < -1$), while the majority of type-I AGNs (15 out of 18) have Eddington ratios above this threshold. For type-I AGNs with reliable estimates of SFRs, the median $\rm log\ \lambda_{Edd}$ is $-0.68$. Thus, we use $\rm log\ \lambda_{Edd} = -0.7$ to divide the type-I AGN sample into high and low Eddington ratio sub-samples, each with 9 objects. In the left panel of Figure~\ref{fig12:Outflow}, we find that type-I AGNs still lie on the linear relation obtained for type-II AGNs. However, the high and low Eddington ratio sub-samples are the same at the outflow strength and sSFRs, with $p-$values of the KS test of two subsamples of the outflow strength and sSFRs being 0.99. This suggests that for our sample the Eddington ratio, i.e., AGN radiation, has a minimal impact on star formation. This might be because the energy and momentum outputs of our sample mainly affect the central regions of the host galaxy, with less influence on star formation in the more distant galactic disk.

In the right panel of Figure~\ref{fig12:Outflow}, we compared the outflow strength and sSFRs of the jetted and other type-I AGNs. We found that the outflow strength and sSFRs of the blue squares still follow the relation of type-II AGNs and the jetted AGNs are located above the red dashed line, which suggests that the radio emission from other type-I AGNs neither promotes nor hinders sSFRs, while the jet activity of jetted AGNs might promote star formation. The $p-$value of the outflow strength of two subsamples is 0.93, and that of sSFRs is 0.72. Although it is not statistically significant, given our small sample size, we consider this to also serve as indirect evidence of feedback. The fact that blue squares are located on the red dashed line also suggests that the mechanisms of radio emissions might be similar, i.e., mostly from star formation.

The enhanced sSFRs of the jetted AGNs do not uniquely imply positive feedback. One alternative explanation is a delayed response between past inflow-driven star formation and the subsequent emergence of AGN-driven outflows or jet activity. In this scenario, cold gas inflows can first trigger star formation, while the radio jet becomes observable after the SMBH accretion activity has developed, naturally placing these sources above the star-forming main sequence at the epoch of observation. Another possibility is that both star formation and jet activity are regulated by a common cold-gas reservoir, such that systems with more abundant gas simultaneously have higher sSFRs and stronger radio emission, without requiring a direct causal link between jets and star formation. Although the $E(B-V)$ values of our jetted sources do not appear systematically higher than those of other type-I AGNs, the limited number of jetted AGNs prevents a robust discrimination between these scenarios. Therefore, the observed sSFR offset should be regarded as suggestive rather than conclusive evidence for jet-driven positive feedback.

\subsection{Radio Luminosity and sSFRs} \label{subsec:SFRRadio}

\begin{figure}[t!]
\includegraphics[width=1.0\linewidth]{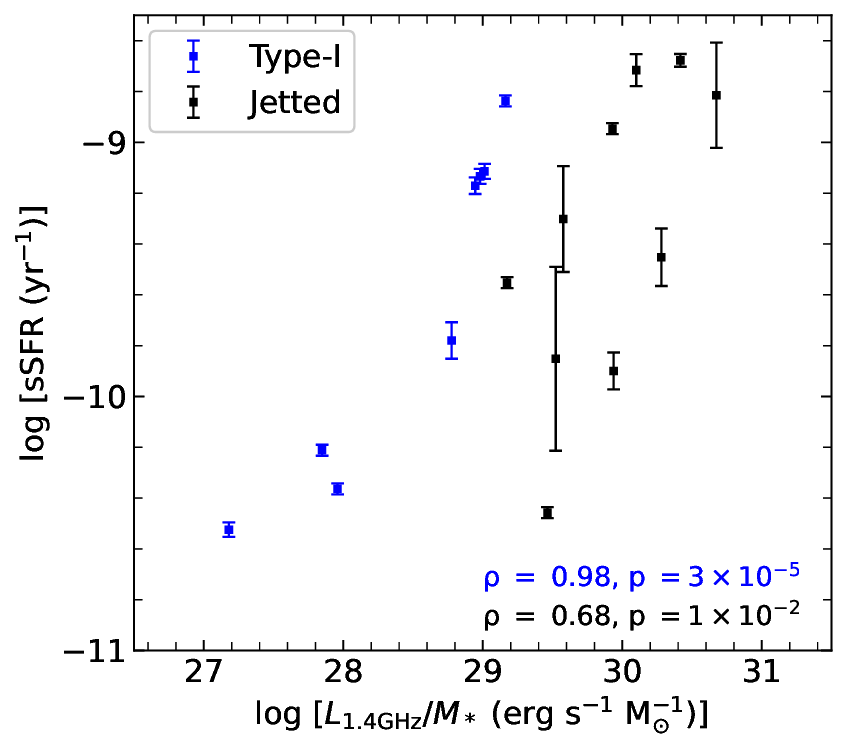}
\caption{Comparison of the radio luminosity divided by stellar mass with the sSFRs for jetted (represented by black squares) and other type-I (indicated by blue squares) AGNs. The Spearman's rank correlation coefficients and $p$-values of each group are shown in the lower-right corners.
\label{fig:Radio_sSFR}}
\end{figure}

To study the possible feedback from jets, in Figure~\ref{fig:Radio_sSFR}, we compare the radio luminosity divided by the stellar mass with the sSFRs of our sample. We divide our sample into jetted (represented by black squares) and other type-I AGNs (indicated by blue squares) as in previous subsections. We found that for both sub-samples, there is a trend where the sSFRs increase with increasing radio emission. For the blue squares, all or part of the radio emission may come from star formation, which results in this positive correlation. Additionally, there might be some selection effects at play because all our sources have radio detections, and sources with larger sSFRs might be more easily selected, leading to this relationship tending towards higher sSFRs and greater radio luminosity. For the black squares, the positive linear correlation between sSFRs and radio luminosity suggests the positive feedback of the jets. Moreover, we also tried to select jetted AGNs using $\rm log\ SFR_{radio} - log\ SFR_{Dust,SED} > 0.5$ and $\rm log\ SFR_{radio} - log\ SFR_{Dust,SED} > 1.5$, and the same correlations still exist, suggesting that this relationship is real.

In addition, for other type-I AGNs, processes unrelated to AGN jets may contribute to the observed trend. In particular, large-scale gas inflows can simultaneously result in enhanced star formation in the host galaxy and induce shocks in the interstellar medium through interactions with AGN-driven winds or turbulent gas motions, leading to enhanced synchrotron radio emission even in the absence of well-collimated jets. Such inflow-related processes naturally produce a positive correlation between sSFR and radio luminosity per stellar mass. Moreover, selection effects associated with requiring radio detections may further bias the sample toward systems with both higher sSFRs and stronger radio emission. For the jetted AGNs, the observed relation may also reflect multiple contributing factors. A common cold-gas reservoir can fuel both star formation and SMBH accretion, giving rise to higher sSFRs and more powerful radio jets without implying a direct causal link between the two.

\section{Summary} \label{sec:6Summary}

We focused on local type-I AGNs ($z<0.3$) exhibiting significant [O~III] outflows, and used the SCUBA-2 on the JCMT to observe 21 AGNs with high radio luminosity, and reduced archival SCUBA-2 data of 21 other objects. Therefore, we constructed a sample of 42 AGNs with a broad radio luminosity range and strong [O~III] outflows for studying the impact of ionized gas outflows and jets on the star formation within their host galaxies, i.e., the feedback of AGNs. The key findings are summarized as follows.

\begin{enumerate}

\item Out of the 42 objects of our sample, 13 were detected at ${\rm 850\ \mu m}$, and 4 at ${\rm 450\ \mu m}$. By analyzing their radio data, we found that the emission of 8 and 1 objects at 850 and 450 ${\rm \mu m}$ is more likely from jet, so their data were replaced with upper limits when fitting SEDs to estimate dust emission.

\item When comparing the ${\rm SFR_{Dust,SED}}$ with the luminosities in the WISE4, IRAS4, and the two SCUBA-2 bands, we discovered the strongest linear correlation with IRAS4. This was followed by WISE4. The luminosities at the SCUBA-2 bands still showed a linear relationship but with larger scatter.

\item We compared the estimated $\rm SFR_{radio}$ and $\rm SFR_{Dust, SED}$, and selected jetted type-I AGNs whose radio emission primarily comes from jet activity.

\item By comparing the SFRs with the main sequence of star formation, we found that jetted type-I AGNs are predominantly above the main sequence. This suggests a general trend of a positive correlation between jet activity and star formation.

\item By comparing the $E(B-V)$ of the objects, and assuming that the gas-to-dust ratio in our sample does not vary from source to source significantly, the jetted type-I AGNs do not have a larger supply of dust and gas. Therefore, the positive correlation between jets and star formation is likely primarily due to feedback.

\item By comparing the dependencies of sSFRs on the Eddington ratio and radio emission within the relationship between sSFRs and outflow strength, we found that the Eddington ratio has no impact on star formation, while jetted type-I AGNs clearly have higher sSFRs. This suggests that the feedback from jets may promote star formation.

\item We compared the radio emission and sSFRs of jetted and other type-I AGNs and found a positive correlation for both sub-samples. We think the correlation for jetted AGNs primarily arises from the positive feedback of the jets, while that for other type-I AGNs may be because their radio emission also has contributions from star formation.

\end{enumerate}

\begin{acknowledgments}

This work has been supported by the National Natural Science Foundation of China (NSFC-12473014, NSFC-12025303, NSFC-12203047), National Key R\&D Program of China No. 2022YFF0503401. This research uses data obtained through the Expanding Partner Program of JCMT (Proposal IDs: M22BP038 and M23AP023).

These observations were obtained by the James Clerk Maxwell Telescope, operated by the East Asian Observatory on behalf of The National Astronomical Observatory of Japan; Academia Sinica Institute of Astronomy and Astrophysics; the Korea Astronomy and Space Science Institute; the National Astronomical Research Institute of Thailand; Center for Astronomical Mega-Science (as well as the National Key R\&D Program of China with No. 2017YFA0402700). Additional funding support is provided by the Science and Technology Facilities Council of the United Kingdom and participating universities and organizations in the United Kingdom and Canada. Additional funds for the construction of SCUBA-2 were provided by the Canada Foundation for Innovation.

The authors wish to recognize and acknowledge the very significant cultural role and reverence that the summit of Maunakea has always had within the indigenous Hawaiian community.  We are most fortunate to have the opportunity to conduct observations from this mountain.

\end{acknowledgments}

\vspace{5mm}
\facilities{GALEX, SDSS, WISE, Herschel(PACS and SPIRE), IRAS, JCMT(SCUBA-2), VLA, GMRT, Planck}

\software{CIGALE \citep{noll_analysis_2009, boquien_cigale_2019, yang_fitting_2022}, BADASS,\citep{sexton_bayesian_2020}, Starlink \citep{currie_starlink_2014}, ORAC-DR \citep{jenness_orac-dr_2015}, SMURF \citep{chapin_scuba-2_2013} }


\appendix

\section{SMBH Properties}\label{A2:MBH}

Accurate measurement of the masses of SMBHs often utilizes the reverberation mapping (RM) method. This technique assumes that light emitted from the accretion disk will be reflected by clouds in the BLR and observed after a time delay due to the finite speed of light. Variability in AGN emission leads to a measurable time delay between the direct observation of accretion disk radiation and its reflected counterpart, allowing for an estimation of the scale of the BLR. The FWHM of broad emission lines represents the velocity dispersion of these clouds. Under the assumption of Keplerian motion, the mass of the SMBH can be estimated.

However, the RM method requires very long observation periods and significant telescope resources. Fortunately, some studies have shown that brighter AGNs have larger BLRs, allowing for the statistical use of AGN luminosity to estimate the scale of the BLR after calibrating by RM results, which is called the single-epoch (SE) method. For our sample, with the FWHM of broad ${\rm H\beta }$ emission lines, we estimate the SMBH mass, as the following formula from \cite{le_calibrating_2020}:
\[
{\rm log}\ M_{\rm BH}\ ({\rm M_{\odot}})=6.87+2.0\times {\rm log\ \frac{FWHM_{H\beta}}{1000\ km\ s^{-1}} +0.53\times log}\ \frac{L_{\rm 5100 \text{\AA}}}{\rm 10^{44}\ erg\ s^{-1}}.
\]
Not all of our objects have a significant broad ${\rm H\beta}$ component. Although there are also some studies \citep{woo_black_2015} about using ${\rm H\alpha}$ lines to estimate the SMBH mass, we still prefer ${\rm H\beta}$ lines. Because our sample is AGNs with strong ionized gas outflows, which might make ${\rm H\alpha}$ broad components mix with outflow components of ${\rm [N\ II]}$ around, and make the FWHM of ${\rm H\alpha}$ overestimated. So given the prominence of the ${\rm H\beta}$ lines within our redshift range and its cleaner spectra, we primarily use the FWHMs of the ${\rm H\beta}$ broad lines, and when ${\rm H\beta}$ broad components are absent, the BLR FWHMs are estimated by using ${\rm H\alpha}$ and the relation in the left panel of Figure~\ref{fig:A3FWHMVOFF}.

\begin{figure*}[ht!]
\includegraphics[width=1.0\linewidth]{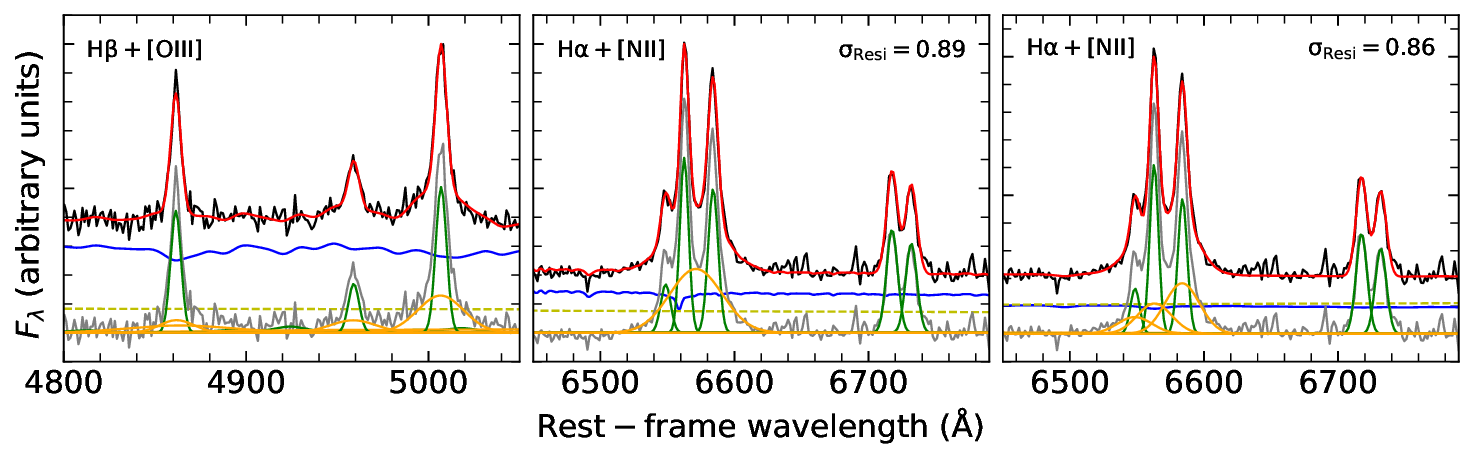}
\caption{Example of decomposition of ${\rm H\beta+[OIII]}$ and the blended ${\rm H\alpha+[NII]}$ for ID22. Black, red, blue, gray, green, and orange curves show observed data, best-fit model, host, emission lines ${\rm(data - host)}$, narrow and broad components. The left panel shows ${\rm H\beta}$ and [O III] lines, while the middle and right panels depict two different approaches to fit the spectra around ${\rm H\alpha}$. In the middle panel, the decomposition utilizes a single broad Gaussian profile. In the right panel, the broad component around ${\rm H\alpha}$ is fitted using three Gaussian profiles, which are constrained to have the same shapes as the [O III] lines shown in the left panel.
\label{fig:5bHaNII}}
\end{figure*}

\begin{figure*}[ht!]
\includegraphics[width=1.0\linewidth]{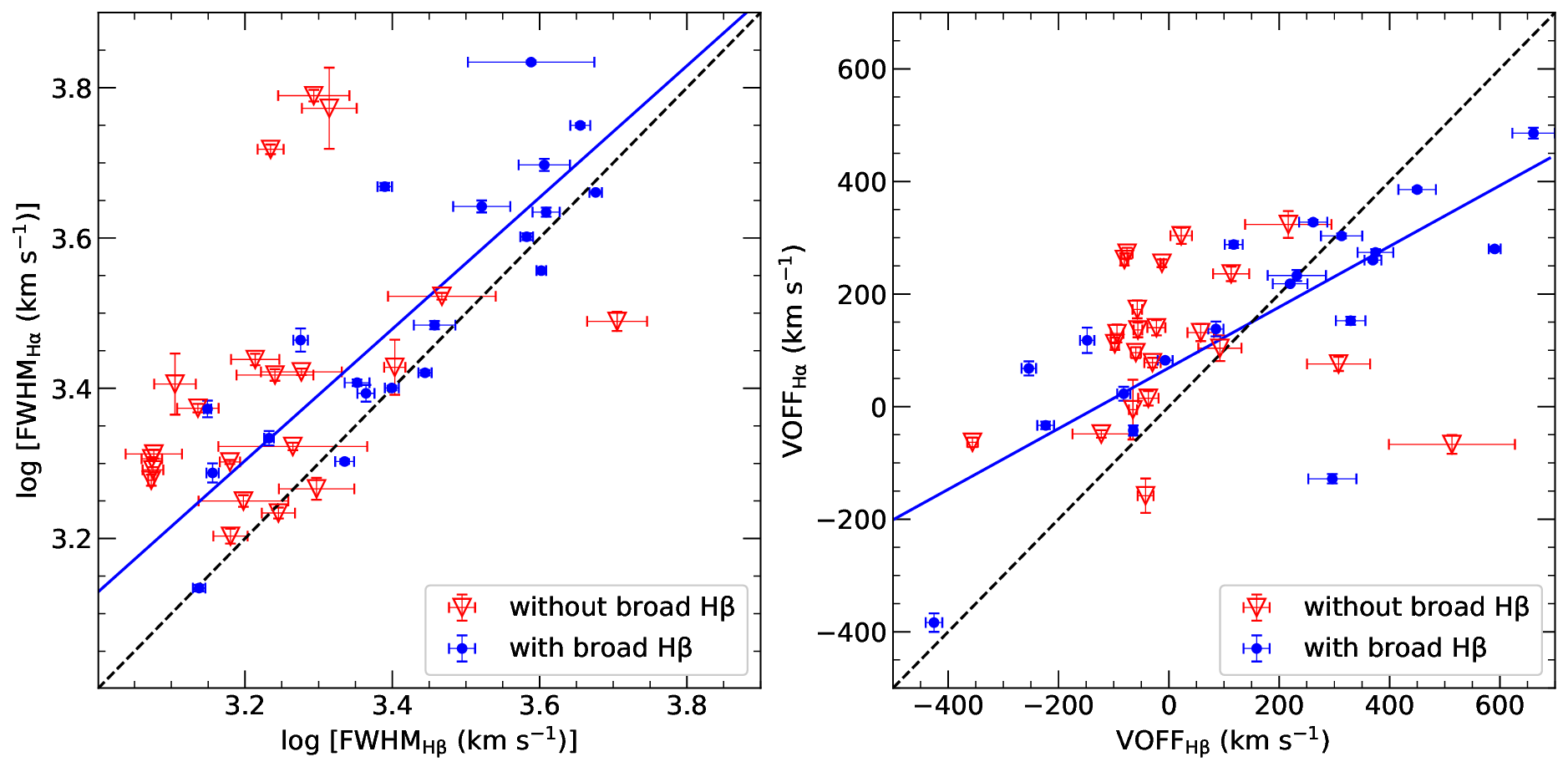}
\caption{Comparison of the FWHMs (left panel) and the first momentum (VOFF, right panel) of ${\rm H\alpha}$ and ${\rm H\beta}$ emission lines. The blue-filled dots and red hallow triangles are targets with and without significant broad ${\rm H\beta}$ components respectively. The black dashed lines show a one-to-one relation, and the blue solid line shows the linear fitting result of blue dots.
\label{fig:A3FWHMVOFF}}
\end{figure*}

\cite{rm_kovabreveceviacutec-rm_dojbrevecinoviacutec_tracing_2022} showed for some type-II AGNs the mixtures of the outflow features of ${\rm H\alpha}$ and [N II] emission lines could be ﬁtted with a single broad Gaussian and be interpreted as ${\rm H\alpha}$ BLR components. Some of our targets also meet the same problem. In Figure~\ref{fig:5bHaNII}, we showed two different ways of the decomposition of the blended ${\rm H\alpha}$+[N II]. The left panel shows no significant broad ${\rm H\beta}$ component, whereas [O III] exhibits strong outflows. In the middle panel, we fit the broad component around the ${\rm H\alpha}$ line with a Gaussian setting all parameter free (amplitude, center wavelength, sigma). In the right panel, we fix the outflow profile of [N II] and ${\rm H\alpha}$ be the same as [O III] and the amplitude ratio of [N II] outflows fixed at 3, leaving only two free parameters (amplitudes of ${\rm H\alpha}$ and [N II]), but achieving a similarly good (or even better) fit. Due to resolution limits, we cannot figure out which one is the truth so when the SMBH masses based on ${\rm H\alpha}$ are considered as the upper limits.

In Figure~\ref{fig:A3FWHMVOFF} we compared the FWHMs (left panel) and the first momentum (VOFF, right panel) of ${\rm H\alpha}$ and ${\rm H\beta}$ emission lines. The blue solid line shows the linear fitting result of blue dots, with the formula as follows:
\[
{\rm log\ FWHM_{H\alpha}\ (km\ s^{-1})=0.88\times log\ FWHM_{H\beta}\ (km\ s^{-1})+0.50},
\]
which is shallower than the slope of \citep{woo_black_2015}. This hints at the contamination of outflow features at ${\rm H\alpha}$ broad component.

Then, similar to \cite{rakshit_census_2018}, the Eddington luminosity was determined using the relation $L_{\rm Edd}=1.26\times 10^{38}M_{\rm BH}$, while the bolometric luminosity was estimated using $L_{\rm bol}=9\times \lambda L_{\rm 5100\text{\AA}}$. Finally, the Eddington ratio was determined by calculating the ratio of bolometric to Eddington luminosity.


\bibliography{sample631}{}
\bibliographystyle{aasjournal}



\end{document}